# Pseudogap with Fermi Arcs and Fermi Pockets in Half-Filled Twisted Transition Metal Dichalcogenides

Yong-Yue Zong,[1] Zhao-Long Gu,[1,*] and Jian-Xin Li[1,2,3,†]

[1]*National Laboratory of Solid State Microstructures and Department of Physics, Nanjing University, Nanjing 210093, China*

[2]*Collaborative Innovation Center of Advanced Microstructures, Nanjing University, Nanjing 210093, China*

[3]*Jiangsu Key Laboratory of Quantum Information Science and Technology, Nanjing University, Suzhou 215163, China*

Twisted transition metal dichalcogenides are a new platform for realizing strongly correlated physics with high tunability. Recent transport experiments [A. Ghiotto *et al.*, Nature (London) **597**, 345 (2021)] have reported the bandwidth-driven evolution of a Mott insulator to a strange metal behavior via the tuning of a displacement field in twisted $WSe_2$ ($tWSe_2$) fixed at half filling. However, the nature of the correlated states and the related Mott physics involved in the whole process remain to be determined. Here, we unveil theoretically the evolution of the ground state of the half-filled moiré Hubbard model as applied to $tWSe_2$, transiting from a pseudogap state with Fermi arcs to a 120° Néel ordered Mott insulator, then to another pseudogap state with Fermi pockets, and eventually to a Fermi liquid via a Lifshitz transition. The pseudogap phases are definitely identified by the vanishing of quasiparticle weights over parts of the Fermi surface, with the remaining parts forming disconnected Fermi arcs or pockets with well-defined quasiparticles. We demonstrate that the Fermi arc or pocket results from the electronic band structure reconstruction driven by electron correlations, marked by the coexistence of the poles and zeros of the single-particle Green's function. We ascribe the ground state of the strange metal featured by the linear-$T$ resistivity observed experimentally in $tWSe_2$ to the second pseudogap state by further calculating the temperature dependence of resistivity. This work reveals the fundamental aspects of the Mottness in moiré system and will stimulate the direct probes of the underlying physics beyond transports via the angle-resolved photoemission spectroscopy and scanning tunneling microscopy.

## I. INTRODUCTION

The fundamental concept of Landau's Fermi liquid theory is the coherent quasiparticle that has a one-to-one correspondence with its noninteracting counterpart. Such correspondence guarantees the stability of the integral Fermi surface despite the presence of interactions between quasiparticles, which leads to a conserved volume enclosed by the Fermi surface known as Luttinger's theorem [1–4]. The existence of quasiparticles at the whole Fermi surface is determined by a finite quasiparticle weight $0 < Z_k < 1$. If the quasiparticle weight $Z_k$ is zero due to the strong correlations between electrons, the system goes beyond the Landau's Fermi liquid paradigm. The exploration of emergent non-Fermi liquid physics constitutes one of the central issues in condensed matter physics.

Pseudogap and strange metal are typical non-Fermi liquid phenomena in strongly correlated systems. These exotic phases are initially observed in high-temperature (high-$T_c$) cuprate superconductors, evolved from its parent antiferromagnetic Mott insulator via dopings [5–26]. The strange metal is characterized by the observation that the resistivity scales as a linear function of temperature ($T$) rather than the characteristic quadratic dependence based on the Landau's Fermi liquid theory of metals. Among the most intriguing phenomena of the pseudogap is the observation of a Fermi surface destruction in the angle-resolved photoemission spectroscopy (ARPES) experiments [5–10], that leads to the formation of disconnected arcs centered on the Brillouin zone (BZ) diagonals. Their understandings are suggested to hold the keys to unveil the mechanism of high-$T_c$ cuprate superconductivity. However, though they are most extensively studied in the framework of high-$T_c$ cuprates, there is still no consensus on their origin. In particular, whether the pseudogap originates solely from the strong correlation effect or from other competing orders is still under debate, as the dopants into the system will spark the competition and cooperation among various orders, thus complicating the understanding of these phenomena [8,23–25]. On the other hand, the Mott insulator–metal transition can also be realized via the bandwidth-controlled manner. Because of the poor ability to continuously tune the bandwidth in conventional quantum materials, studies of the bandwidth-controlled transitions are rare [27–33]. Thanks to the recent advance in moiré systems, twisted transition metal dichalcogenides (tTMDs) provide a highly tunable platform to realize Mott insulator–metal transition by continuously tuning the bandwidth via a displacement field [34–41]. Intriguingly, the transport experiment on twisted $WSe_2$ ($tWSe_2$) has discovered strange metal behavior in proximity to the Mott transition marked by the linear-$T$ resistivity [36], which is similar to that observed in high-$T_c$ cuprates. It is noted that this strange metal phase can be realized through the change of displacement field, while the system is fixed at half filling. Thus, no external competing order is expected to be induced. Now, the study of strong correlation effects in tTMDs is still in its infancy. It calls for theoretical investigations of the underlying Mott physics and the advancement of the direct probes of the properties of the Fermi surface and quasiparticles beyond transports, such as the nano-ARPES and the scanning tunneling microscopy (STM) experiments.

*Contact author: waltergu@nju.edu.cn

†Contact author: jxli@nju.edu.cn

Motivated by these perspectives, in this work, we study the nature of the correlated states and the related non-Fermi liquid physics involved in the half-filled $tWSe_2$ based on the moiré Hubbard model. Through a comprehensive analysis of the interacting band structures, Fermi surfaces, momentum distribution functions, and quasiparticle weight, we unveil that the ground state undergoes bandwidth-controlled pseudogap–Mott insulator–pseudogap–Fermi liquid transitions solely tuned by a displacement field. In these two pseudogap phases, the Fermi surfaces are destructed into separated Fermi arcs and Fermi pockets, respectively, and the arc or pocket regions change dramatically with the displacement field although the system is fixed at half filling. Such facts indicate that these pseudogap phases, as proximate states of the Mott insulator, are non-Fermi liquids resulting from intrinsic strong correlations, which is further elaborated by the coexistence of poles and zeros of the single-particle Green's function $G$ [42–46]. The shapes and distributions of the Fermi arcs and pockets, and the resulting quasiparticle interference patterns, are unequivocally demonstrated. Moreover, the pseudogap with Fermi pockets at zero temperature is elaborated to be responsible for the strange metal behavior observed experimentally, indicating that the strange metal is the finite-temperature continuation of the pseudogap in this system. This work elaborates that twisted transition metal dichalcogenides are promising platforms for studying bandwidth-controlled Mottness and will stimulate future experimental investigations beyond transports.

We summarize the underlying picture for the pseudogap phases with Fermi arc or pocket based on our theoretical results, in comparison to those for the Mott insulator and the weakly correlated systems (Fermi liquid, band insulator, and Slater insulator) in Fig. 1. Let us start from the case of the Mott insulator, in which the fully gapped upper and lower Hubbard bands are linked by zeros of $G$ ($G = 0$, the white dashed lines) and the Luttinger surface determined by zeros separates the BZ into $\mathrm{Re}G > 0$ and $\mathrm{Re}G < 0$ regions [see Fig. 1(a)]. In the case of Fermi arc, the energy band in the destructed region ($k$ direction) is also split and linked by zeros of $G$ [see Fig. 1(b)], whereas in the arc region ($k'$ direction) the energy band remains as an ungapped segment determined by the poles of $G$ ($G = \infty$, the yellow solid lines). While in the case of Fermi pocket the energy bands in both the destructed and pocket region are split and connected by zeros [see Fig. 1(b)]. This point shares the similarity with the Mott insulator. However, in the pocket region, the chemical potential $\mu$ does not cross the band gap, but the lower split band. Since the top part of the split band is bent down, $\mu$ crosses this band two times and forms a pocket. So both pseudogaps are characterized by the coexistence of poles and zeros of the single-particle Green's function $G$. The zero-connecting energy band splitting demonstrates that it results from the Mott physics. Contrastively, in weakly correlated systems such as Fermi liquids, band insulators, and Slater insulators [see Fig. 1(c)], only $G = \infty$ exists and neither the shift of chemical potential nor the band folding arising from antiferromagnetic ordering can result in the emergence of zeros of $G$. We thus ascribe these pseudogap phases to the bridge states between the Mott insulator and the Fermi liquid.

Our work is structured as follows. We introduce the moiré Hubbard model for $tWSe_2$ and discuss its phase diagram in Sec. II. We unveil the characteristic features of the pseudogap phases with interacting density of states (DOS) and Fermi surfaces, and pin them down by the quasiparticle weights obtained from momentum distribution functions along different paths in BZ in Sec. III. We elaborate the pseudogap phases as non-Fermi liquids resulted from strong correlations by the coexistence of poles and zeros of $G$, and give a comprehensive picture for Mott insulators, pseudogap states, and Fermi liquids based on interacting band structures in Sec. IV. Proposals for experimental tests of the Fermi arcs and pockets in tTMDs are presented in Sec. V. The relevance of the pseudogap to the experimentally observed strange metal is discussed in Sec. VI. We conclude and provide a discussion in Sec. VII.

## II. MODEL AND PHASE DIAGRAM

We adopt the single-orbital triangular lattice moiré Hubbard model, which is believed to capture the main

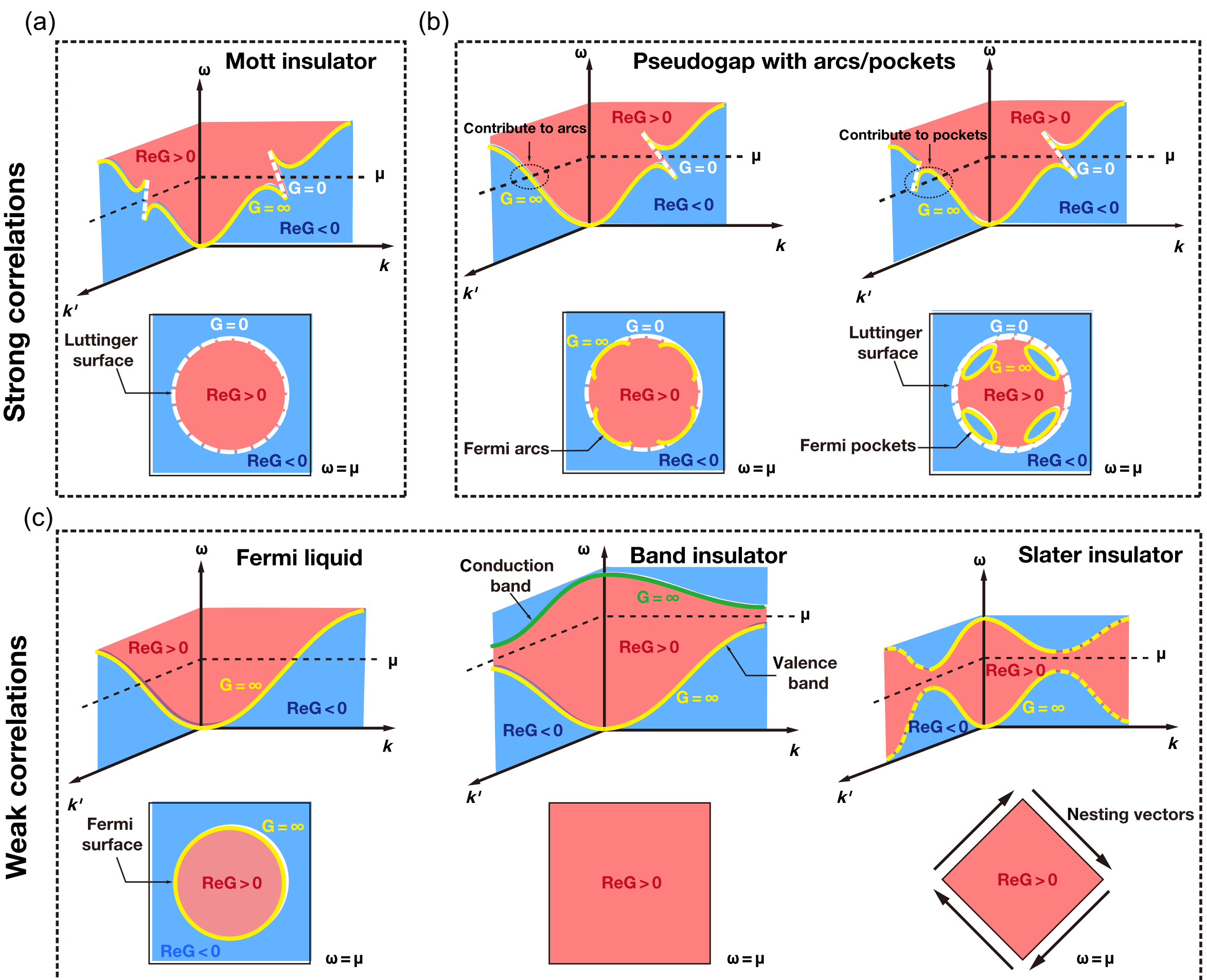

FIG. 1. A comprehensive illustration of band structures characterized by poles and zeros of Green's function ($G$). (a) Mott insulator. (b) Pseudogap with Fermi arcs or pockets. (c) Weakly correlated systems such as Fermi liquid, band insulator, and Slater insulator. For weakly correlated systems, including Landau Fermi liquid, band insulator, and the Slater insulator caused by band folding arising from antiferromagnetic ordering, there are no zeros of $G$ ($G = 0$, white dashed line). The quasiparticle dispersion $\varepsilon_k$ is determined by poles of $G$ ($G = \infty$, yellow line), and the real part of $G$ (Re$G$) changes sign across these poles. A Fermi liquid has a Fermi surface (the yellow circle, at $\mu$) determined only by poles, which divides the Brillouin zone (BZ) into Re$G > 0$ (red) and Re$G < 0$ (blue) regions with the former satisfying the Luttinger's theorem. The difference of the band insulator and Slater insulator compared to Landau Fermi liquid is the lack of a Fermi surface and Re$G > 0$ at the whole BZ at $\mu$, while in a Mott insulator, the interactions split the band into the upper and lower Hubbard bands (UHB and LHB). The zeros of $G$ emerge as connecting lines between the ends of UHB and LHB. At $\mu$, a Mott insulator has a Luttinger surface determined only by zeros, which also divides the BZ into Re$G > 0$ and Re$G < 0$ regions, in stark contrast to the band insulator and Slater insulator. The pseudogap is a "composition" of Fermi liquid and Mott insulator from the perspective of band structures. At $\mu$, it hosts zeros in some directions of the momentum space and poles in the others. The pseudogap with Fermi arcs has ungapped segments of the band constituting the arcs at $\mu$ and gapped segments linked by zeros. The pseudogap with Fermi pockets has split bands linked by zeros, but the lower band crosses $\mu$ two times forming the pockets.

physics of $tWSe_2$ [34–36,41,47,48]. The Hamiltonian is expressed as

$$H = \sum_{i,j,\sigma} t^{\sigma}_{ij} c^{\dagger}_{i,\sigma} c_{j,\sigma} + U \sum_{i} n_{i\uparrow} n_{i\downarrow}, \qquad (1)$$

where $i$, $j$ represent the first to the fourth nearest-neighbor sites on the triangular superlattice, and $U$ is the on-site repulsive interaction. The spin-dependent hoppings $t^{\sigma}_{ij}$ are shown in Fig. 2(a), which rely on the displacement field $D$ and are obtained from the tight-binding approximation (TBA) of the topmost moiré band in the continuum model (CM) [49]. The parameters in CM are determined by fitting the first-principle results of $tWSe_2$ in Ref. [34]. We have considered up to the fourth nearest-neighbor hoppings to obtain a good agreement between the TBA and CM energy bands (see Appendix A for the band structures by the CM and TBA). The final results of the hoppings as a function of $V^z$ are shown in Fig. 2(b). Here, $V^z$ is the layer-dependent potential representing $D$ in the experiment [36] and has an approximate linear dependency on $D$ (see Appendix A). The interacting single-particle Green's function of the moiré Hubbard model Eq. (1) is calculated based on a 12-site cluster tiling of the triangular lattice by use of the cluster perturbation theory (CPT) [50] and the variational cluster approach (VCA) [51], both of which have been successfully applied to quantum many-body systems [52–60]. The selection of the 12-site cluster balances the largest size compatible with our computational resources and preservation of the full point group symmetry of the system. On other clusters that do not maintain these symmetries, we could find qualitatively similar results but with explicitly broken threefold rotation symmetry. The details of CM, TBA, CPT, and VCA are provided in Appendixes A and B. In this work, we focus on the twisted angle $\theta = 4.2°$ in Ref. [36], where the bandwidth-controlled metal–Mott insulator–metal transitions and strange metal behavior near the phase boundary have been reported.

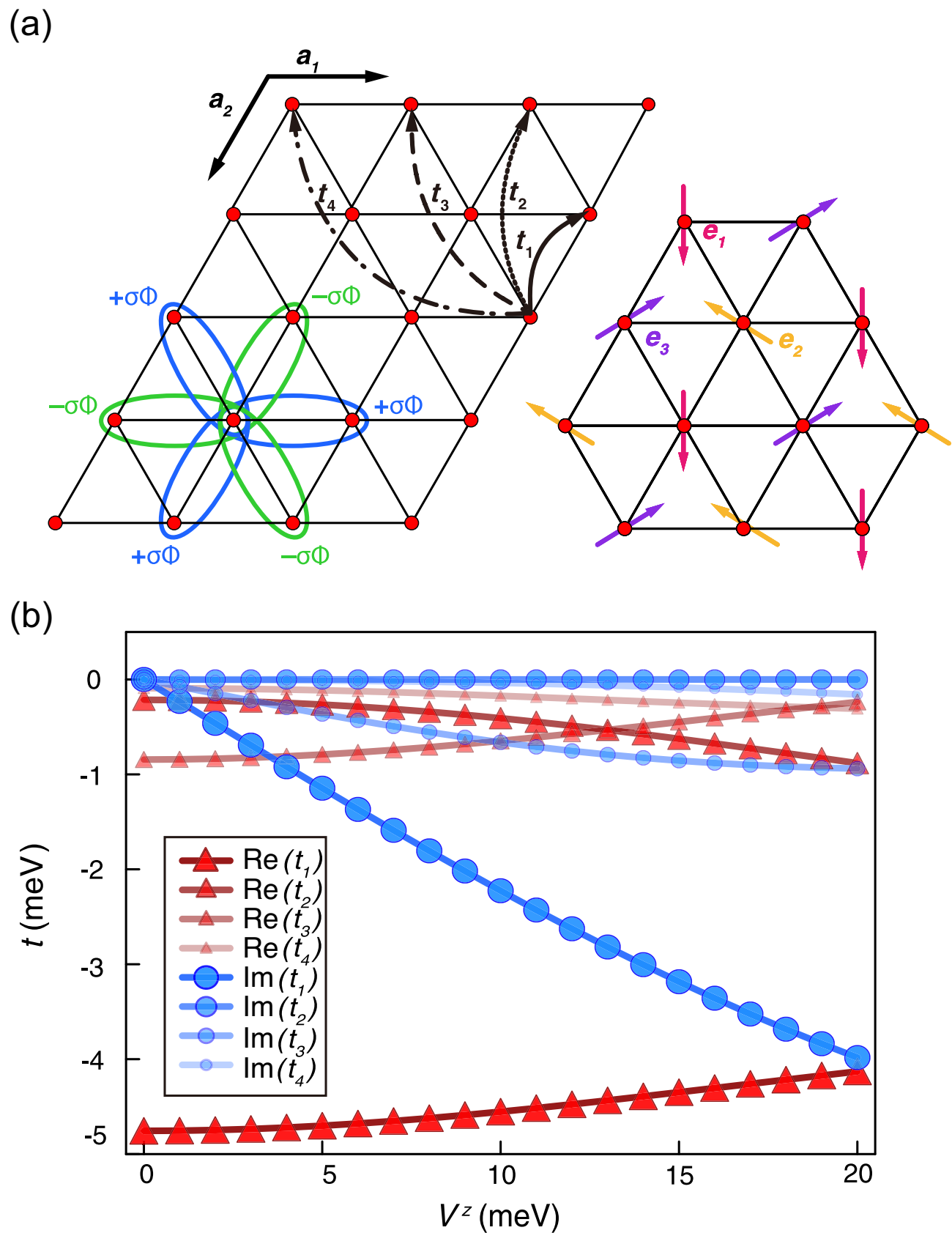

FIG. 2. (a) Illustration of the hoppings in the triangular lattice moiré Hubbard model (left) and 12-site cluster [47] of the triangle lattice in quantum cluster computations (right). Here, $t_1$, $t_2$, $t_3$, $t_4$ are the first to fourth nearest-neighbor hoppings. $\pm\sigma\phi$ denotes the phases for the hoppings of spin-up ($\sigma = +1$) and spin-down ($\sigma = -1$) electrons. $e_1$, $e_2$, $e_3$ show the directions of 120° Néel order. (b) $t_1$, $t_2$, $t_3$, $t_4$ as a function of $V^z$, where the real part and imaginary part are denoted by the red triangular and blue circular dots, respectively. The values of these parameters also can be found in Table I.

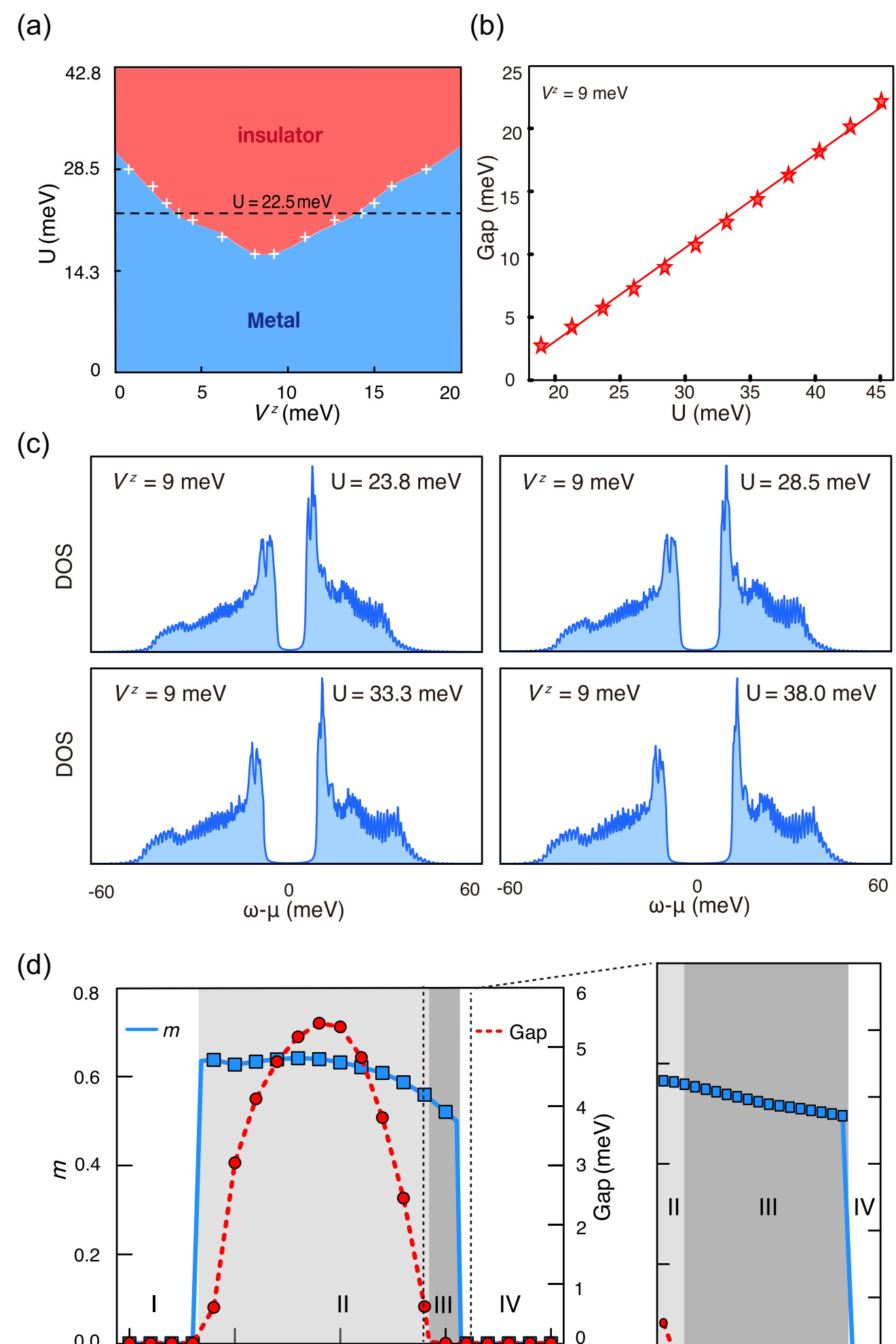

FIG. 3. (a) $U$-$V^z$ phase diagram of $tWSe_2$ at half filling. The red and blue regions represent the insulating phase and metallic phase, respectively. (b) Single-particle gap as a function of $U$ at $V^z = 9$ meV. (c) DOS that are used to map (b). (d) Single-particle gap (red) and 120° Néel order parameter $m$ (blue) as functions of $V^z$ at $U = 22.5$ meV. The regions labeled by I, II, III, and IV represent the paramagnetic pseudgap phase, the 120° Néel ordered Mott insulating phase, the 120° Néel ordered pseudogap phase, and the normal paramagnetic metal, respectively.

In Fig. 3(a), we present the $U$-$V^z$ phase diagram of the ground state of Eq. (1) at half filling. We map this phase diagram by measuring the single-particle gap from the density of states. The extracted gap as a function of $U$ and some typical DOS at $V^z = 9$ meV are shown in Figs. 3(b) and 3(c), respectively. We find that the phase diagram is divided into two regimes, namely, the metal without gap at the chemical potential $\mu$ and the insulator with a full gap at $\mu$. In the whole regime of $V^z$, for $0 < U < 16$ meV and $U > 29$ meV, the system perpetually resides in the metallic and insulating phases, respectively. Within $16 < U < 29$ meV, the system experiences transitions from a metallic to an insulating phase then to another metallic phase with $V^z$, which is qualitatively consistent with the previous results in $tWSe_2$ at half filling [36], where the second metal-insulator transition was observed experimentally while the first metal-insulator transition was inferred from their Hartree-Fock calculation. We select a typical value of $U = 22.5$ meV to present our results in this article unless otherwise specified. Other choices of $U$ around $U = 22.5$ meV will not change the results qualitatively.

Then, we show the single-particle gap as a function of $V^z$ by the red dashed line in Fig. 3(d). It can be seen that an energy gap is opened in the range of $4 < V^z < 14.2$ meV and the maximum value of the gap is about 5.5 meV, which is in good agreement with the experimental results [see Fig. 1(e) in Ref. [36] ]. In Fig. 3(d), we also show the moment of the 120° Néel order $m$ [the orientations are shown as $e_1$, $e_2$, $e_3$ in Fig. 2(a)] as the blue line, which is consistent with the previous studies [41,48]. With a combination of $m$ and the single-particle gap, we can conclude that the first metallic state is paramagnetic and the intermediate insulating phase is 120° Néel ordered, while the second metallic state is magnetically ordered in a small range of $14.2 < V^z < 15.8$ meV and is paramagnetic when $V^z > 15.8$ meV, which can be seen more clearly in the magnified inset of the small window ($14 < V^z < 16$ meV) in Fig. 3(d).

## III. PSEUDOGAP, FERMI ARCS, FERMI POCKETS, AND COHERENT QUASIPARTICLES

The typical DOS of the four phases with $V^z = 3, 9, 15$, and 17 meV are presented in Fig. 4(a) and the corresponding noninteracting DOS are shown as the red dashed lines. For the intermediate insulating phase [Fig. 4(a), $V^z = 9$ meV], a full gap is seen clearly. We find that this gap increases linearly with $U$ [see Fig. 3(b)], so we identify this intermediate insulating phase as a Mott insulator. For the second paramagnetic metallic phase [Fig. 4(a), $V^z = 17$ meV], the DOS follows the same behavior as the noninteracting DOS, only with a slight renormalization due to the interaction effect. Remarkably, for both the first paramagnetic metallic phase [Fig. 4(a), $V^z = 3$ meV] and the magnetically ordered metallic phase [Fig. 4(a), $V^z = 15$ meV], the depressions of the DOS appear but residual DOS remains at the Fermi level $\mu$. This kind of phenomenon with partially opened gap is the so-called pseudogap as firstly named in high-$T_c$ cuprates [10,42,61–65].

To further investigate the emergence of the pseudogap, we show the single-particle spectral function $A(\boldsymbol{k},\omega) = -\mathrm{Im}\,G(\boldsymbol{k},\omega)/\pi$ at $\omega = \mu$ in the first BZ with the same typical $V^z$ values in Fig. 4(b). Because of the explicit SU(2) spin rotation symmetry breaking resulted from the spin-dependent hoppings in Eq. (1), we also show the spin-down component of $A(\boldsymbol{k},\omega = \mu)$ in Fig. 4(c). Its spin-up component can be obtained by the inversion of the spin-down one centered on $\Gamma$ point. In Figs. 4(b) and 4(c), the noninteracting Fermi surfaces are also plotted as the red dotted lines for comparison.

For the Mott insulator [Figs. 4(b) and 4(c), $V^z = 9$ meV], the noninteracting Fermi surfaces are fully gapped by $U$, thereby no spectral weight is observed at $\mu$ exhibiting as a whole dark region. For the second paramagnetic metal [Figs. 4(b) and 4(c), $V^z = 17$ meV], the interacting Fermi surfaces coincide with the noninteracting ones exactly, indicating the fulfillment of Luttinger's theorem. Therefore, this metallic phase is a Landau's Fermi liquid as the interaction does not change the Fermi surface [66]. Significantly, for the pseudogap phases [Figs. 4(b) and 4(c), $V^z = 3$ and 15 meV], the Fermi surfaces are destructed in some portions of the momentum space, while the remaining "bright" parts of $A(\boldsymbol{k},\omega = \mu)$ in other portions constitute Fermi arcs and Fermi pockets. The destructed portions account for the depression of the DOS at $\mu$ and the Fermi arcs and pockets contribute to the residual DOS in the pseudogap states, which results in the partially opened pseudogaps as shown in the DOS plots in Fig. 4(a) with $V^z = 3$ and 15 meV. Such a small Fermi surface in the pseudogap phases evidently violates the quasiparticle's one-to-one correspondence to its noninteracting counterpart, suggesting these pseudogap phases are non-Fermi liquids that go beyond the Landau Fermi paradigm.

To pin down this point, we go further to calculate the momentum distribution function $n(\boldsymbol{k}) = \int_{-\infty}^{\mu} A(\boldsymbol{k},\omega)d\omega$ and show the results of spin-down electrons along three typical high-symmetric paths in the first BZ in Fig. 4(d) (the spin-up electrons are equivalent to the spin-down electrons as long as we change the $K$ to $K'$ and $K'$ to $K$). The corresponding noninteracting $n(\boldsymbol{k})$ are shown as the dashed red lines in Fig. 4(d) as well. For Landau's Fermi liquid phase [Fig. 4(d), $V^z = 17$ meV], the clear discontinuity marked by double-sided arrows on $n(\boldsymbol{k})$ exists in all the paths that cross the Fermi surface (e.g., the $\Gamma$–$K$ and $\Gamma$–$M$ paths) in the first BZ. On the other hand, the paths that do not cross the Fermi surface (e.g., the $\Gamma$–$K'$ path) correspond to fully occupied states [$n(\boldsymbol{k}) \approx 1$], indicating that this path lies entirely within the Fermi surface. The nonzero step (discontinuity) verifies the existence of the quasiparticle, as the height of the step is equal to the quasiparticle weight $Z_k$ representing the overlap between the noninteracting particle and the interaction-dressed quasiparticle [67]. Our results also show the exact coincidence of the position of discontinuity with the noninteracting $n(\mathbf{k})$, reflecting perfectly the analytical properties of the self-energy in a Fermi liquid $\mathrm{Re}\Sigma(\boldsymbol{k},\omega) \propto (\boldsymbol{k} - \boldsymbol{k}_F)$. The effect of interactions in this phase is to renormalize $Z_k$ with $0 < Z_k < 1$. The existence of nonzero $Z_k$ on all the paths that cross the Fermi surface guarantees the one-to-one correspondence in this Fermi liquid phase. For the Mott insulating phase [Fig. 4(d), $V^z = 9$ meV], along all paths, $n(\boldsymbol{k})$ decreases continuously across the noninteracting Fermi surface, indicating $Z_k = 0$. Thus, coherent quasiparticles completely disappear in this phase. The textbook-standard numerical result for the momentum distribution function $n(\boldsymbol{k})$ obtained for the Fermi liquid and Mott insulator benchmarks the significance of our theoretical approach to the studies of quasiparticle properties, and it enables us to pin down the existences and shapes of the Fermi arcs and pockets. As can be seen in Fig. 4(d) with $V^z = 3$ meV and $V^z = 15$ meV, for the pseudogap phases, the step of $n(\boldsymbol{k})$ exists in the paths across the "bright" portions of $A(\boldsymbol{k}, \omega = \mu)$ (e.g., the $\Gamma$–$K$ path), while the step disappears in the paths through the destructed portions (e.g., the $\Gamma$–$M$, $\Gamma$–$K'$ paths). This confirms unambiguously the absence of coherent quasiparticles in partial portions of the momentum space and thus the breakdown of the one-to-one correspondence. We note that, for the first pseudogap phase ($V^z = 3$ meV), although $A(\boldsymbol{k}, \omega = \mu)$ seems to exhibit a bright region along the $\Gamma$–$M$ path, $n(\boldsymbol{k})$ definitely exclude the existence of coherent quasiparticles here. On the other hand, for the second pseudogap phase, $A(\boldsymbol{k}, \omega = \mu)$ is dim at the outer edges of the pockets along the $\Gamma$–$K$ path; however, $n(\boldsymbol{k})$ accurately verify the existence of quasiparticles by a small step in $n(\boldsymbol{k})$. Therefore, compared to the ambiguity based on the intensity of $A(\boldsymbol{k}, \omega = \mu)$, we can rigorously identify the start and end points of Fermi arcs and the edges of Fermi pockets by $n(\boldsymbol{k})$. Additionally, the positions of the steps in both pseudogap phases are slightly shifted by interactions in comparison to the noninteracting cases, which is also different from the Landau's Fermi liquid phase.

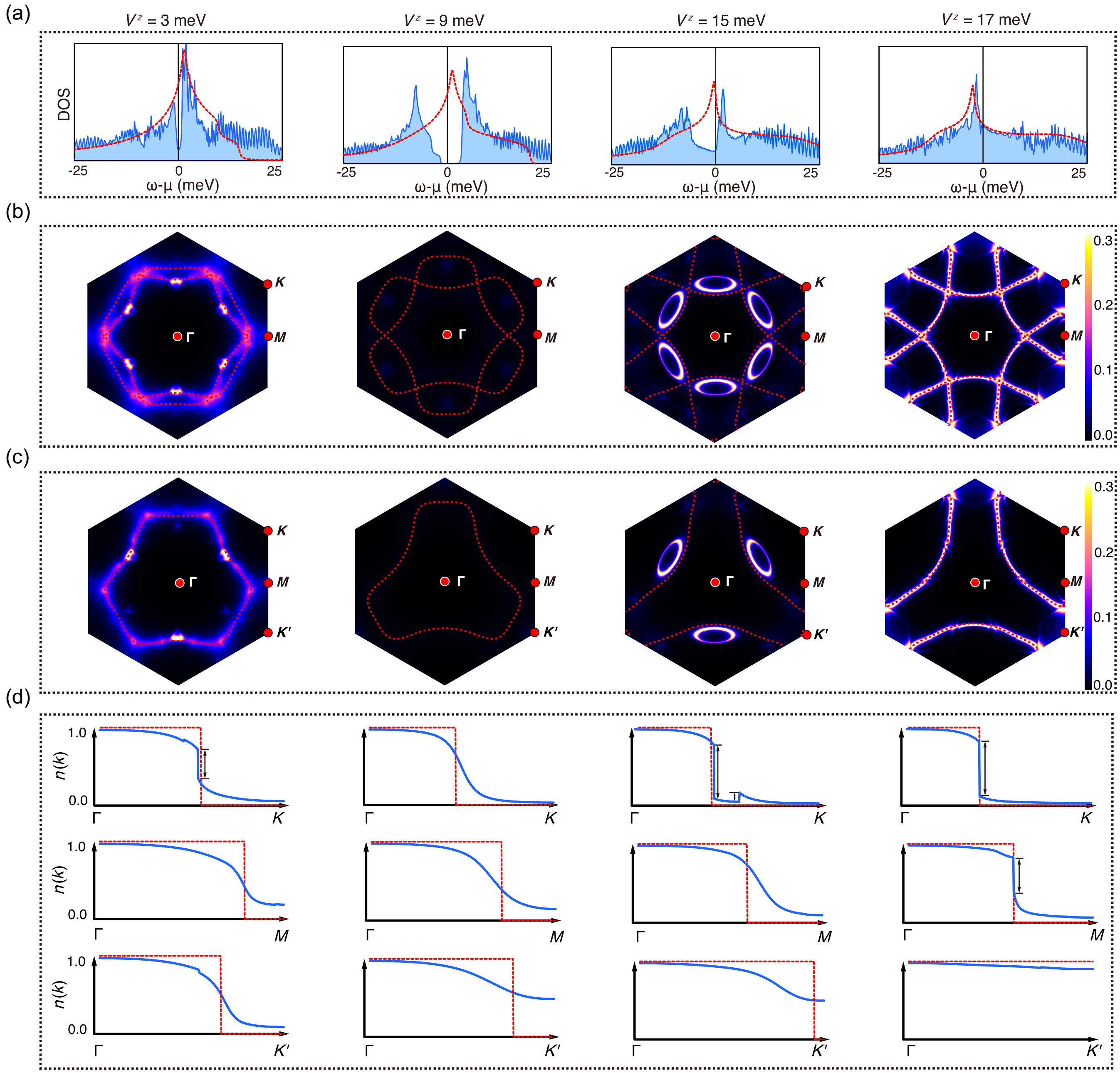

FIG. 4. Density of states (DOS), spectral function, and momentum distribution function $n(k)$ at $V^z = 3$, 9, 15, and 17 meV. (a) DOS for noninteracting (red dotted lines) and interacting (blue regions) systems. (b) Spectral function at the Fermi level $\mu$ for interacting systems in the first Brillouin zone. Red dotted lines denote the noninteracting Fermi surfaces. (c) Spin-down component of (b). (d) $n(\boldsymbol{k})$ of spin-down electrons along the $\Gamma$–$K$ (first row), $\Gamma$–$M$ (second row), and $\Gamma$–$K'$ (third row) paths. Here, the blue solid lines and the red dotted lines denote the results of interacting and noninteracting systems, respectively.

We can also understand the distributions of the Fermi arcs and pockets in the first BZ via the numerical results for the quasiparticle weight $Z_k$. This quantity is related to the renormalized effective mass $m^*$ as $Z_k \sim m/m^*$, where $m$ is the bare mass. We find that $Z_k$ in the $\Gamma$–$M$ path is lesser than that in the $\Gamma$–$K$ path in the Landau's Fermi liquid phase as shown in the Fig. 4(d) with $V^z = 17$ meV. Therefore, the quasiparticle has a larger $m^*$ in the $\Gamma$–$M$ path, indicating that the interaction has a more significant impact on the quasiparticle along this path. Considering the shape of the full Fermi surface in the Fermi liquid phase, we call the region in the $\boldsymbol{k}$ space around the $\Gamma$–$K$ path as the diagonal region, and the other the off-diagonal region. Our numerical results indicate that the quasiparticles situated in the off-diagonal region are most strongly scattered by the Hubbard interaction, while those in the diagonal region are lesser. As a result, when entering into the pseudogap phases, the Fermi surface in the off-diagonal region is destructed by this strong scattering effect on its quasiparticles, while that in the diagonal region remains and exhibits either arc or pocket depending on the energy band reconstructions. This shares some similarity with the situation in high-$T_c$ cuprates [8,24], where the diagonal (node) direction is along $\Gamma = (0, 0)$ to $M = (\pi, \pi)$ in the square BZ and the antinode direction is along $\Gamma$ to $X = (0, \pi)$. In the pseudogap phase there, the antinode region is destructed while the diagonal region remains as an arc [8,24].

## IV. UNDERLYING PHYSICS REVEALED BY POLES AND ZEROS OF GREEN'S FUNCTION

From the above discussions, it can be inferred that the pseudogap phase is the bridge state between the Mott insulator and the Landau's Fermi liquid. In view of this perspective, we present the real part of the Green's function $G(\boldsymbol{k}, \omega = \mu)$ for the spin-down component in Fig. 5, with the same parameters as those in Fig. 4. Here, the white lines correspond to the zeros of $G$ ($G = 0$) and the discontinuity lines from yellow to blue correspond to the poles of $G$ ($G = \pm\infty$).

For the Mott insulator [Fig. 5(a), $V^z = 9$ meV], only zeros of $G$ are seen near the Fermi level, which enclose a connected surface, namely, the Luttinger surface [44,68,69]. For the Fermi liquid [Fig. 5(a), $V^z = 17$ meV], only poles of $G$ are found, which defines the Fermi surface satisfying the Luttinger's theorem. Strikingly, for both pseudogap phases [Fig. 5(a), $V^z = 3$, 15 meV], there is a coexistence of poles and zeros of $G$. Specifically, for the first pseudogap phase with Fermi arcs [Fig. 5(a), $V^z = 3$ meV], the poles of $G$ exist on the arcs and the end points of the separated arcs are linked by zeros of $G$, while for the second pseudogap phase with Fermi pockets [Fig. 5(a), $V^z = 15$ meV], the poles of $G$ locate at the pockets constituting the small Fermi surfaces and the zeros of $G$ enclose the Luttinger surface surrounding

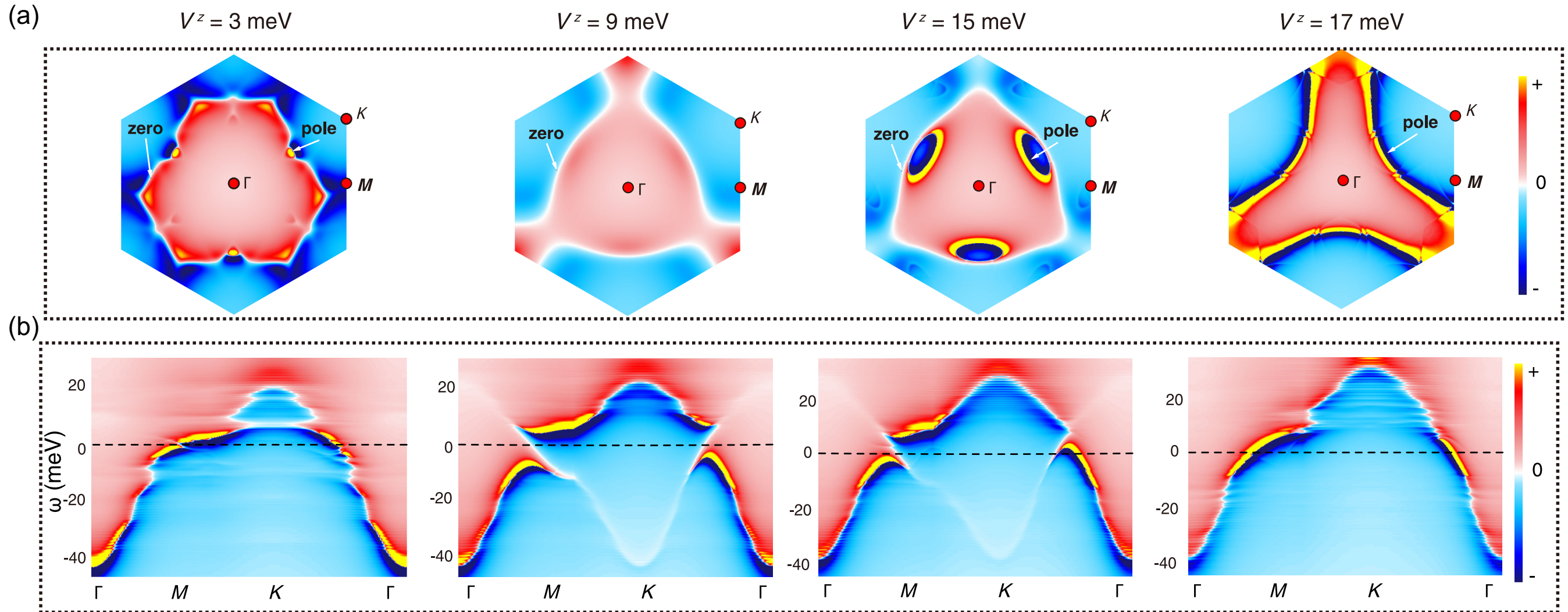

FIG. 5. Poles and zeros of Green's function at $V^z = 3$, 9, 15, and 17 meV. Panels (a) and (b) show the intensity maps of the real part of Green's function Re$G$ (spin-down component) at Fermi level and along the $\Gamma$–$M$–$K$–$\Gamma$ path, respectively. The poles of $G$ are shown as the discontinuity lines from Re$G = +\infty$ (blue) to Re $G = -\infty$ (yellow), and the zeros of $G$ are shown as Re$G = 0$ (white).

all the separated pockets. These numerical results are also demonstrated schematically in Fig. 1.

To understand the evolution of the zeros and poles of the Green's function $G$, let us look at the general form of $G$ in the presence of interactions,

$$G(\boldsymbol{k},\omega)=\frac{1}{\omega-\varepsilon(\boldsymbol{k})-\Sigma(\boldsymbol{k},\omega)}, \tag{2}$$

where $\varepsilon$ is the dispersion of noninteracting quasiparticles and $\Sigma$ is the self-energy resulting from interactions. For a Fermi liquid, $\Sigma$ converges to zero as $\omega$ approaches $\mu$; therefore, there are only poles ($\omega=\varepsilon(\boldsymbol{k})$) but no zeros of $G$ [e.g., Fig. 5(a) at $V^z=17$ meV]. As mentioned above, the property $\Sigma(\boldsymbol{k},\omega=\mu)=0$ also explains that the Fermi surface does not change regardless of the interactions in the Fermi liquid phase. For a Mott insulator, $\Sigma$ diverges ($\Sigma=\infty$) at $\mu$; consequently, there are only zeros of $G$ but no poles [e.g., Fig. 5(a) at $V^z=9$ meV]. The absence of poles is consistent with the system hosting a charge energy gap. The existence of zeros demonstrates that the generation of the gap should be attributed to the Mott mechanism rather than the Slater mechanism although the ground state hosts 120° magnetic order. This is because, for a Slater insulator, the system can be modeled by a static mean-field Hamiltonian [70] so that $\Sigma$ will never diverge (the difference of the band structure between the Mott and Slater insulators can be seen schematically in Fig. 1 as well). For the pseudogap phases [e.g., Fig. 5(a) at $V^z=3$, 15 meV], $\Sigma$ possesses strong $\boldsymbol{k}$ dependence: it diverges at some $\boldsymbol{k}$ points resulting in the destructed portions (i.e., the $G=0$ regions) in the band structures, while it gets finite values at other $\boldsymbol{k}$ points leading to a reconstruction of the band structures and thus the emergence of Fermi arcs or pockets (i.e., the $G=\pm\infty$ regions). The reconstruction of the band structures obtained by our numerical calculations of Eq. (1) is shown in Fig. 5(b), which is reproduced with the schematic diagram in Fig. 1.

Despite the fact that antiferromagnetism is favored in the pseudogap phase with Fermi pockets, the structure of the poles and zeros of $G$ does not change when the magnetic order is excluded by force in quantum cluster calculations. In Fig. 6, we show the Fermi surface, momentum distribution function and Re$G$ for the spin-down component without 120° Néel order, and find that the structures of Fermi pockets and Luttinger surfaces do not change qualitatively compared with the results shown in Figs. 4 and 5. This proves that the emergence of Fermi pockets is absolutely not caused by the folding of Fermi surfaces. To conclude, the pseudogap found here is the bridge state between the Mott insulator and the Landau's Fermi liquid, in that the destructed portions are determined by the zeros of $G$ and the arc or pocket portions are determined by the poles of $G$.

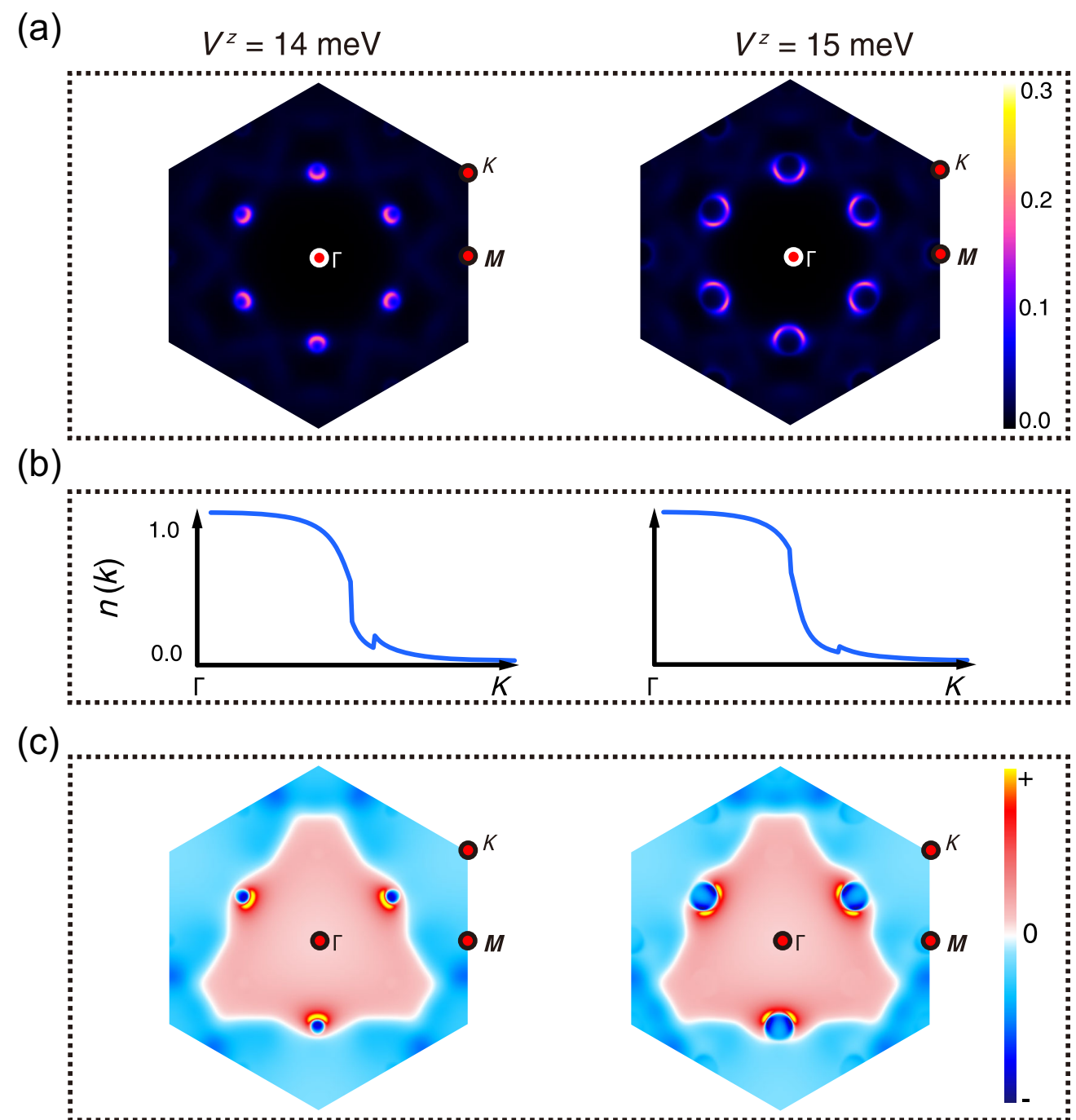

FIG. 6. Pseudogap phase with Fermi pockets without magnetic order. (a) Fermi surfaces, (b) momentum distribution function, and (c) Re$G$ for the spin-down component.

## V. PROPOSAL FOR EXPERIMENTAL TEST OF FERMI ARCS AND POCKETS

The position and shape of the Fermi arc and pocket in the Brillouin zone can serve as unique features for further experimental probes of the pseudogap states. In this regard, we present the evolution of Fermi arcs and Fermi pockets of spin-down electrons in the pseudogap phases in Fig. 7. The Fermi arc starts to emerge from $V^z=0$. It exhibits a sixfold symmetric arc distribution, as the SU(2) spin rotation symmetry is preserved in this case. When increasing $V^z$, along with the breaking of the SU(2) symmetry, the distribution of the arc evolves from the sixfold to the threefold rotation symmetry. In this evolution progress, three of the arcs gradually shrink their arc length and lose their spectral intensity with $V^z$. On the other hand, for the pseudogap phase with Fermi pockets, the volume of the Fermi pocket expands continuously with $V^z$ and finally turn into the large Fermi surface in the Fermi liquid phase via a Lifshitz transition. Such unveiled features of the Fermi arcs and Fermi pockets can be probed directly by future (spin-polarized) nano-ARPES experiments.

In addition to the momentum distribution curves, ARPES experiments can also probe the pseudogap behavior via energy distribution curves (EDCs). For better comparison with experiments, we provide EDCs for the pseudogap phases with Fermi arcs and Fermi pockets. In Fig. 8(a), the peaks of EDCs cross the Fermi level once, where the zero-frequency peak represents the bright segment of the Fermi arc. By contrast, in Fig. 8(b), the peaks of EDCs cross the Fermi level twice, exhibiting two distinct zero-frequency peaks with the second peak noticeably suppressed, which unambiguously mark the inner and outer edges of the Fermi pocket. In Figs. 8(c) and 8(d), which plot EDCs through the destructed portion of the Fermi surfaces, one clearly observes a strong suppression of spectral weight at $\omega = \mu$ and the emergence of nearly coherent peaks on either side, including the characteristic backbending behavior of the peaks as they approach the Fermi level.

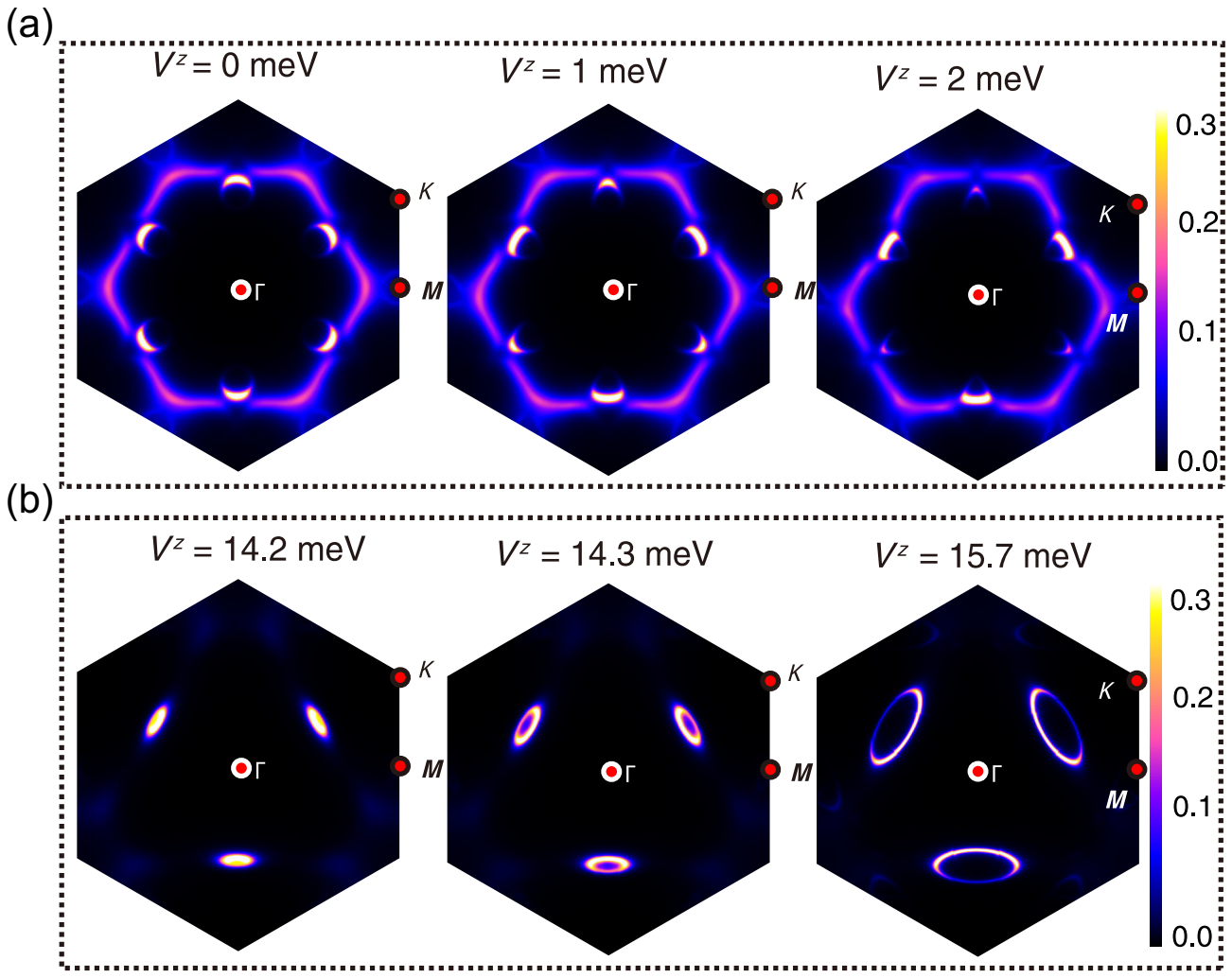

FIG. 7. Evolution of Fermi arcs (a) and Fermi pockets (b) in the pseudogap phases with $V^z$. Only the spin-down part of Fermi surfaces are shown here.

Moreover, the existence of Fermi pockets could be verified further by measurements of the de Haas–van Alphen effect and the Shubnikov–de Haas effect. Both experiments can detect quantum oscillations, indicating two-dimensional pockets for electrons.

Another route to identify these theoretical predictions is to measure the quasiparticle scattering interference (QPI) patterns in the two pseudogap phases with the STM experiments [71–73]. The QPI occurs between the ingoing and scattered outgoing electrons by impurity, where the joint density of states $\rho_q(\omega)$ of the initial and final states forms spatial modulation which can be visualized directly by the STM experimental results after a Fourier transformation. $\rho_q(\omega)$ is written as

$$\rho_q(\omega) = -\frac{1}{\pi}\mathrm{Im}\sum_{k\sigma}\mathrm{Tr}(G_{k\sigma}T_{k\sigma,k+q\sigma}G_{k+q\sigma}), \tag{3}$$

where $T_{k\sigma,k+q\sigma} = V_{k,k+q} + \sum_{k'} V_{kk'}G_{k'\sigma}T_{k'\sigma,k'+q\sigma}$ is the scattering $T$ matrix, $G_{k\sigma}$ is the single-particle Green's function, and $V_{kk'}$ is the scattering potential. As usually used [71–73], we adopt a $\delta$-type scattering potential $V_{k,k+q} = V_0$. In the energy window around the Fermi level, we calculate $\rho_q(\omega)$ of the pseudogap states with Fermi arcs and Fermi pockets and show these QPI patterns in Fig. 9. In Fig. 9(a), the region and intensity of $\rho_q(\omega)$ grow rapidly as $\omega$ leaving Fermi level. It is in agreement with the band structure of the pseudogap phase with Fermi arcs in Figs. 1(b) and 5(b) ($V^z = 3$ meV), where $\omega$ crosses the energy band around the $M$ point as soon as it slightly deviates from the Fermi level that results in the DOS increasing sharply and enhances the chances of scatters between electrons. In Fig. 9(b), the closed circles in QPI patterns are formed due to the scattering of electrons within and between the Fermi pockets, where the shrinking of these circles with the increase of $\omega$ reflects the shrinking of Fermi pockets with the increase of $\omega$ as seen in Figs. 1(b) and 5(b) ($V^z = 15$ meV). The above QPI patterns provide a guide to be compared with the future STM experiments on tWSe2 to identify the Fermi arc and Fermi pockets.

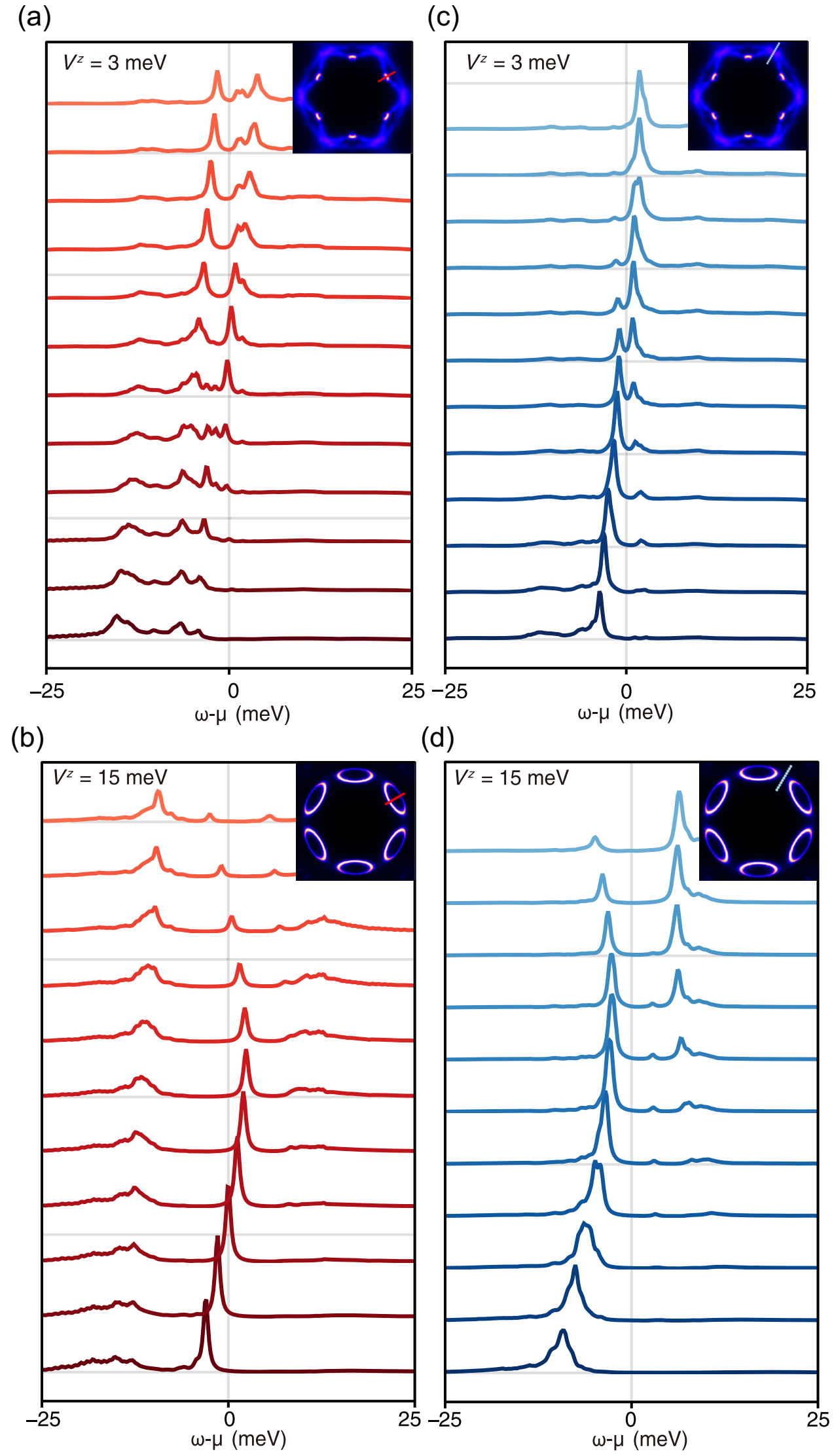

FIG. 8. Energy distribution curves (EDCs) of the pseudogap phase with Fermi arcs and Fermi pockets, taken along the momentum space paths shown in the insets. The curves from bottom to top correspond to the points on the paths from the inner end toward the outer end. Panels (a) and (b) follow the path across the Fermi arc and the Fermi pocket, respectively. Panels (c) and (d) go through the destructed portions of the Fermi surfaces.

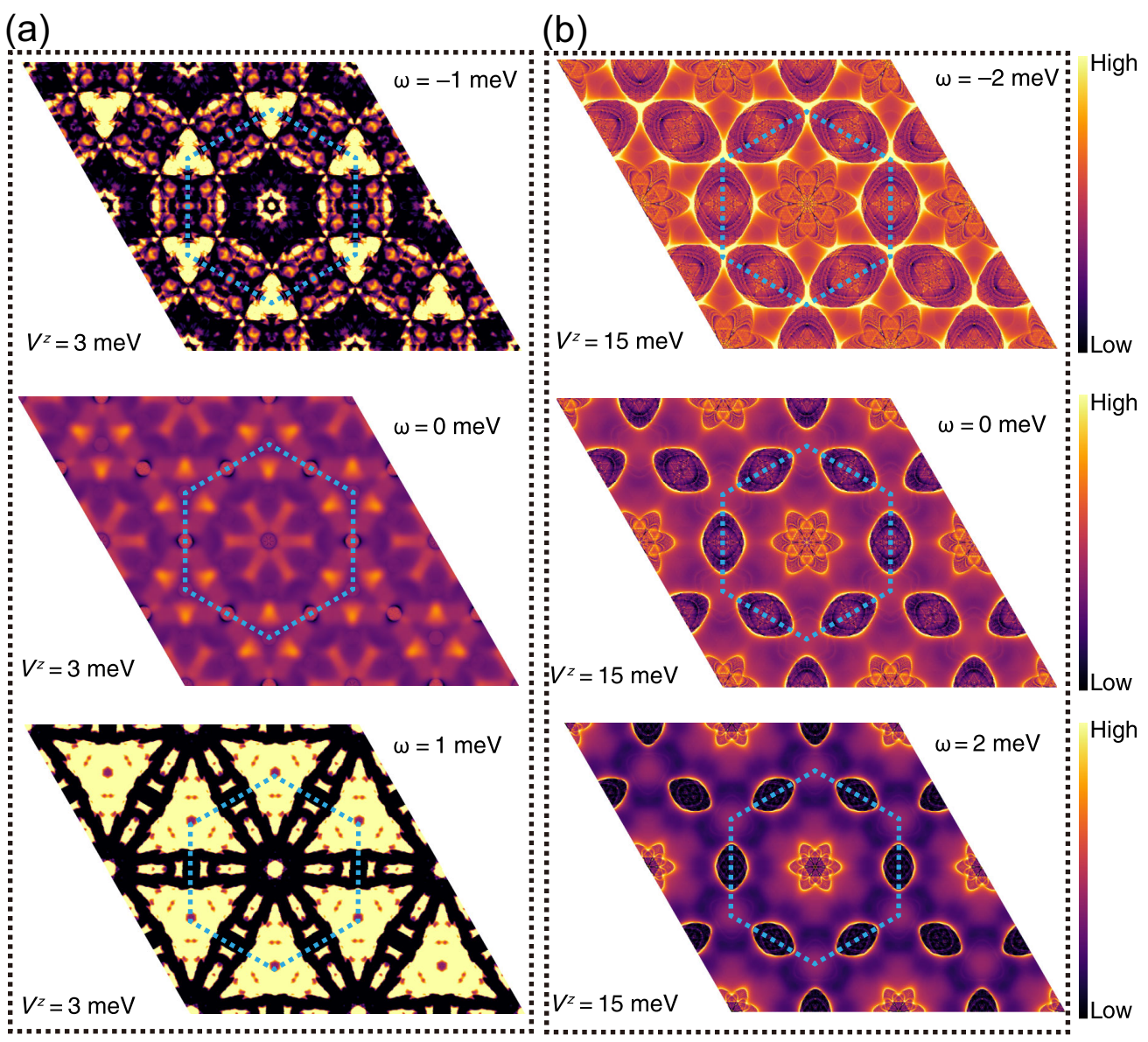

FIG. 9. (a) Quasiparticle interference (QPI) patterns of pseudogap states with Fermi arcs at $\omega = -1$, 0, and 1 meV. (b) QPI patterns of pseudogap states with Fermi pockets at $\omega = -2$, 0, and 2 meV.

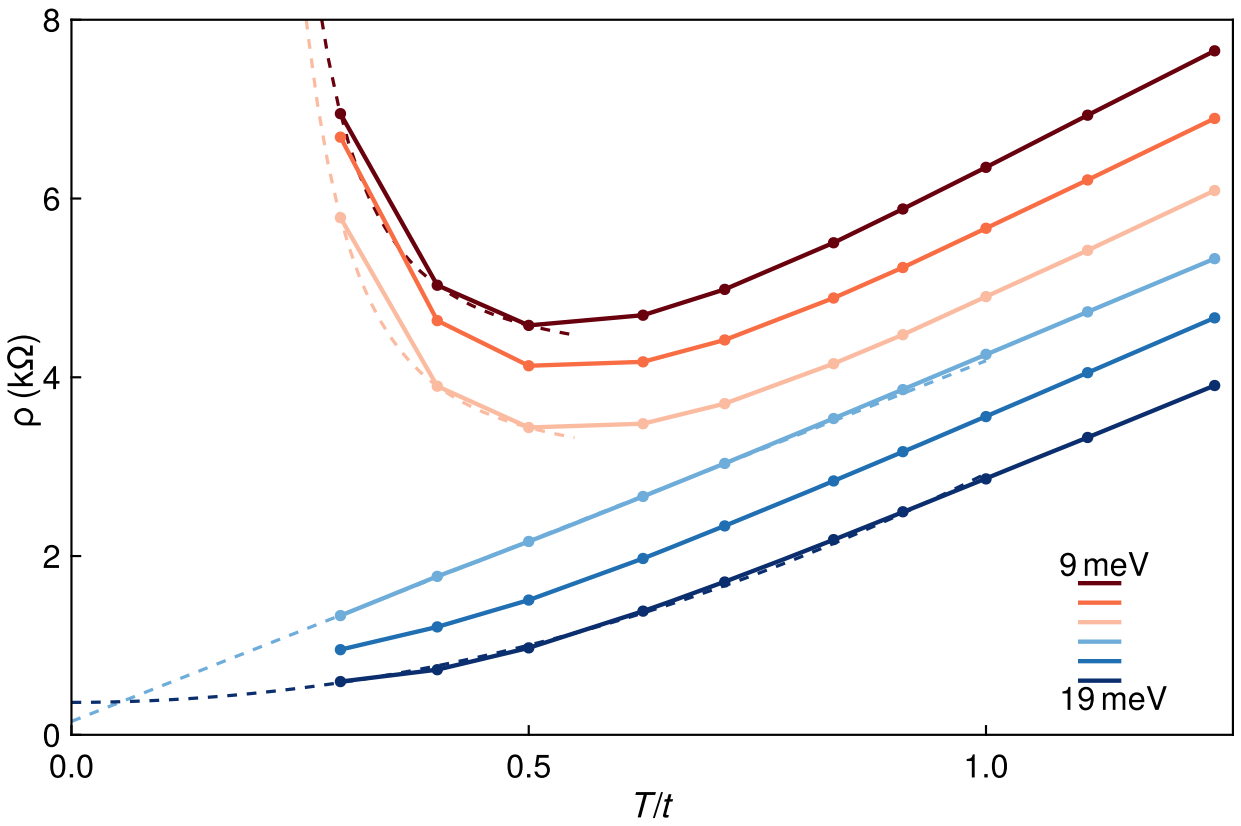

FIG. 10. Resistivity versus temperature for the displacement field $V^z = 9$, 11, 13, 15, 17, 19 meV, revealing the insulator–strange metal–Fermi liquid transitions. The system starts with an insulating behavior characterized by $\rho = A_1 e^{B_1/T}$ at low temperatures, transitions to a strange metal with linear-$T$ resistivity $\rho = A_2 T + B_2$, and ultimately evolves into a Fermi liquid phase following $\rho = A_3 T^2 + B_3$ at low temperatures. The dashed lines are guides for the eye.

## VI. RELEVANCE OF THE PSEUDOGAP TO THE EXPERIMENTALLY OBSERVED STRANGE METAL

The transport experiment on $tWSe_2$ has discovered a strange metal behavior in proximity to the Mott transition marked by the linear-$T$ resistivity [36]. We will show that this kind of non-Fermi liquid behavior at low temperatures is related to the pseudogap phase at zero temperature. To this end, we calculate the conductivity of the model Eq. (1) in the process of the Mott transition at finite temperatures by the following formula [41,74],

$$\sigma_{\mathrm{dc}}(\beta) \propto \sum_{\boldsymbol{k}} \left( \frac{\partial \varepsilon_{\boldsymbol{k}}}{\partial k_x} \right)^2 \int d\omega A_\beta(\boldsymbol{k}, \omega)^2 \frac{-\partial n_F(\omega)}{\partial \omega}, \quad (4)$$

where $\varepsilon_{\boldsymbol{k}}$ is the noninteracting energy band, $A_\beta(\boldsymbol{k}, \omega)$ is the single-particle spectral function at temperature $T = 1/\beta$, and $n_F(\omega)$ is the Fermi-Dirac distribution function.

To obtain the spectral function $A_\beta(\boldsymbol{k}, \omega)$ at finite temperatures, we change the cluster solver of CPT from exact diagonalization (ED) to time-dependent variational principle (TDVP) to perform imaginary- and real-time evolutions. The detail of the method can be found in Appendix C. In Fig. 10, we show the resistivity $\rho$ ($\rho = 1/\sigma_{\mathrm{dc}}$) versus temperature with increasing the displacement field from $V^z = 9$ to 19 meV. The temperature dependence of $\rho$ evolves from exponential at low temperatures with $V^z = 9$, 11, and 13 meV, to a $T$ linear with $V^z = 15$ meV, and finally to a $T^2$ scaling at low temperatures with $V^z = 17$ and 19 meV. It indicates the transition from an insulating phase to a strange metal and then to a Fermi liquid phase, which is consistent with the experimental result [36]. Significantly, such phase transitions at finite temperatures coincide with the Mott insulator–pseudogap–Fermi liquid transitions identified at zero temperature above. This consistency strongly suggests that the strange metal phase represents the finite-temperature continuation of the pseudogap state in this system. Hence, we ascribe the ground state of the strange metal featured by the linear-$T$ resistivity observed experimentally in $tWSe_2$ [36] to the pseudogap state with Fermi pockets here.

We also cross-check the behaviors of resistivity by calculating the current-current correlation function on an $8 \times 8$ lattice with determinant quantum Monte Carlo (DQMC) simulations presented in Appendix D. The DQMC results as shown in Fig. 13 are consistent qualitatively with the TDVP-based CPT results in Fig. 10.

## VII. CONCLUSION AND DISCUSSION

Based on the moiré Hubbard model describing the physics of $tWSe_2$ at half filling, we numerically evidence the emergence of pseudogap states with Fermi arcs and pockets in proximity to the Mott insulator, by a comprehensive analysis of the density of states, quasiparticle weight, momentum distribution function, and the poles and zeros of the interacting single-particle Green's function. Moreover, we elaborate that the pseudogap phase with Fermi pockets at zero temperature is responsible for the strange metal behavior observed experimentally, by calculating the resistivity with the finite-temperature Green's function.

Our results show confirmedly that a Hubbard model can produce the pseudogap phases with Fermi arcs or Fermi pockets via the electron energy band reconstruction which realizes the bandwidth-controlled Mott transition. The pseudogap phase with arcs is a paramagnetic state, while that with pockets is the 120° magnetic ordered state. We note that the Fermi pockets is an intrinsic phenomenon rather than a replica of a folded Fermi arc induced by magnetic ordering based on the following two reasons: (i) the underlying mechanism for the emergence of pseudogaps (Fig. 1) has been elaborated by the coexistence of zeros and poles of the Green's function; and (ii) the Fermi pockets persist in the CPT calculations even in the absence of magnetic order [where CPT does not allow for spontaneous symmetry breaking (Fig. 6)]. From the energy band structure illustrated in Fig. 1(b) and calculated in Fig. 5(b), the Fermi pocket arises from the two-time crossing of the chemical potential through the bended band. The bending of the energy band results from the gap opening due to the Hubbard interaction, which is slightly above the chemical potential, while in the case of the Fermi arc, the energy band in the arc region is nothing but the renormalized one of the bare band, so the chemical potential crosses the band only once. These properties indicate that the pseudogap with Fermi pockets is physically closer to the Mott insulator. The concurrence of the Fermi arc and Fermi pocket in a single model system is striking, as it is the only example as far as we know. In the most extensively studied high-$T_c$ cuprates, Fermi arc in the pseudogap phase has been observed commonly via ARPES measurements [5–10]. Only recently, a small nodal pocket was unambiguously observed in the deeply hole-underdoped inner $CuO_2$ planes in a five-layer overall-hole-doped cuprate [75], where a long-range antiferromagnetic order exists. However, the experiment shows that the pseudogap does not develop in the inner planes, which is ascribed to the lack of electrons required to generate the gapped excitations in the band structure. The Hubbard model we studied is fixed at half filling, so half of the Brillouin zone is enclosed by the Fermi surface, when the ratio of the energy band to the Hubbard $U$ is large and the system is at a metal phase. Once the system evolves into the pseudogap phases, only arcs or pockets remain, the electron in the gapped region loses its quasiparticle attribute with a zero weight. Thus, the pseudogap states with Fermi arcs and pockets we unveiled are non-Fermi liquid states. Our theoretical results will serve as a guide for further experimental and theoretical studies of the Mott physics in the Mott insulator–metal transition, in particular that controlled by the bandwidth.

## ACKNOWLEDGMENTS

We thank L. Wang for helpful discussions. The work was supported by National Key Projects for Research and Development of China (Grants No. 2021YFA1400400 and No. 2024YFA1408104), the National Natural Science Foundation of China (Grants No. 12550405, No. 92165205, No. 12434005, and No. 12404174), and Natural Science Foundation of Jiangsu province (Grant No. BK20230765). We thank e-Science Center of Collaborative Innovation Center of Advanced Microstructures for support in allocation of CPU.

## DATA AVAILABILITY

The data that support the findings of this article are openly available [76].

## APPENDIX A: TIGHT-BINDING APPROXIMATION

For small and incommensurate twist angles, the first-principle calculations for moiré systems are difficult to perform. In such cases, the continuum model introduced by Refs. [48,49] well describes the valence bands at $\pm K$ valleys. Because of the spin-valley locking resulting from the strong spin-orbital coupling, in $+K$ valley, only the spin-up valence band needs to be considered and its Hamiltonian is as follows:

$$H_\uparrow = \begin{pmatrix} -\frac{\hbar^2(\boldsymbol{k}-\boldsymbol{\kappa}_+)^2}{2m} + V^z + \Delta_+(\boldsymbol{r}) & \Delta_T(\boldsymbol{r}) \\ \Delta_T^\dagger(\boldsymbol{r}) & -\frac{\hbar^2(\boldsymbol{k}-\boldsymbol{\kappa}_-)^2}{2m} - V^z + \Delta_-(\boldsymbol{r}) \end{pmatrix}, \tag{A1}$$

where

$$\Delta_\pm(\boldsymbol{r}) = 2V \sum_{j=1,3,5} \cos(\boldsymbol{b}_j \cdot \boldsymbol{r} \pm \psi),$$
$$\Delta_T(\boldsymbol{r}) = w(1 + e^{-i\boldsymbol{b}_2\cdot\boldsymbol{r}} + e^{-i\boldsymbol{b}_3\cdot\boldsymbol{r}})$$

and

$$m = 0.45 m_0, \qquad \kappa_\pm = \frac{4\pi}{3a_M}\left(-\frac{\sqrt{3}}{2}, \mp\frac{1}{2}\right),$$
$$a_M = \frac{a_0}{2\sin(\theta/2)}, \qquad a_0 = 3.30 \text{ Å}.$$

Here, $m_0$, $a_0$, $\theta$, $\boldsymbol{b}_j$ are the electron rest mass, monolayer lattice constant, twist angle, and moiré reciprocal lattice vectors in the first shell, respectively. $\boldsymbol{b}_j$ ($j = 2, 3, \ldots, 6$) are related to $\boldsymbol{b}_1 = [4\pi/(\sqrt{3}a_M)](1, 0)$ by $(j-1)\pi/3$ rotation. The Hamiltonian of the spin-down valence band in $-K$ valley is the time-reversal partner of Eq. (A1). The phenomenological parameters $(V, \psi, w)$ could be estimated using the first-principle results at commensurate twist angles. In this work, we find $(V, \psi, w) \approx (-1.28 \text{ meV}, 22.7°, -12.9 \text{ meV})$ by fitting the energy bands of Eq. (A1) to those obtained by the first-principle calculations for the twisted $WSe_2$ at $\theta = 5.08°$ [34]. The layer-dependent potential $V^z$ is a tunable parameter resulting from the nonzero vertical displacement field $D$ in experiments, which has an approximate linear dependency on $D$: $V^z/(\mathrm{meV}) \approx 47.1 \times D/(\mathrm{V/nm})$.

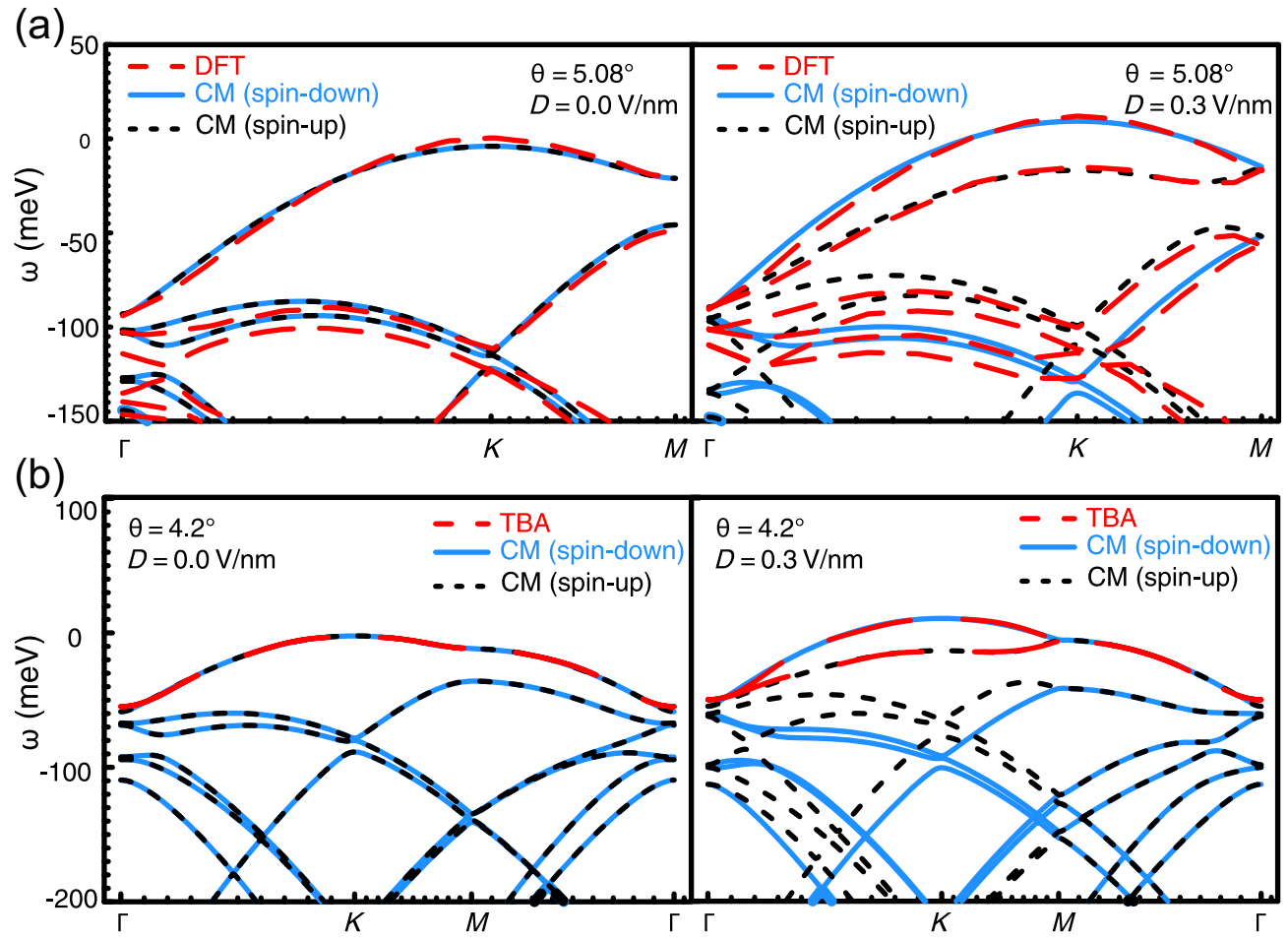

FIG. 11. Band structures in the continuum model (CM) and tight-binding approximation (TBA). (a) Band structures of $tWSe_2$ at $\theta = 5.08°$ with $D = 0$ and $0.3\mathrm{V/nm}$, respectively. The red dashed lines are the DFT results extracted from Ref. [34]. The blue lines and black dotted lines denote the spin-down part and spin-up part, respectively, by use of CM to fit the DFT results. (b) Band structures of $tWSe_2$ at $\theta = 4.2°$ with $D = 0$ and $0.3\mathrm{V/nm}$, respectively. The blue lines and black dotted lines denote the spin-down part and spin-up part, respectively, computed with CM. The red dashed lines are the topmost moiré band calculated via TBA.

The continuum model Eq. (A1) can be extrapolated to the incommensurate twist angle $\theta = 4.2°$, which is realized in the $tWSe_2$ experiment [36]. In the whole displacement field $D$ regime in the experiment, the effective tight-binding model for this band is defined on a triangular lattice [34,36,48]. The Hamiltonian reads

$$H_{\mathrm{TB}} = \sum_{i,j,\sigma} t^{\sigma}_{ij} c^{\dagger}_{i\sigma} c_{j\sigma} + \mathrm{H.c.}, \qquad t^{\sigma}_{ij} = \frac{1}{N} \sum_{\boldsymbol{k} \in M} e^{i\boldsymbol{k}\cdot\boldsymbol{r}_{ij}} \varepsilon_{\sigma}(\boldsymbol{k}), \tag{A2}$$

where $\boldsymbol{r}_{ij} = \boldsymbol{R}_i - \boldsymbol{R}_j$ are triangular moiré superlattice vectors, $\varepsilon_{\sigma}(\boldsymbol{k})$ are the eigenvalues of the topmost moiré band in CM and $M$ represents the first moiré Brillouin zone. The results for the continuum model and tight-binding approximation are shown in Fig. 11 and Table I.

## APPENDIX B: CLUSTER PERTURBATION THEORY

In order to discuss the strong correlations effects at half filling of the $tWSe_2$, the Hubbard interaction is considered along with Eq. (A2), and the moiré Hubbard model is written as

$$H = H_{\mathrm{TB}} + H_U = \sum_{i,j,\sigma} t^{\sigma}_{ij} c^{\dagger}_{i\sigma} c_{j\sigma} + \mathrm{H.c.} + U \sum_{i} n_{i\uparrow} n_{i\downarrow}. \tag{B1}$$

TABLE I. Hopping parameters with respect to $V^z$. All values are in meV.

| $V^z$ | $t_1$.re | $t_1$.im | $t_2$.re | $t_2$.im | $t_3$.re | $t_3$.im | $t_4$.re | $t_4$.im |
|---|---|---|---|---|---|---|---|---|
| 0 | −4.755 | 0.0 | −0.215 | 0.0 | −0.844 | 0.0 | −0.092 | 0.0 |
| 1 | −4.753 | −0.231 | −0.217 | 0.0 | −0.842 | −0.075 | −0.093 | −0.002 |
| 2 | −4.747 | −0.462 | −0.223 | 0.0 | −0.835 | −0.149 | −0.095 | −0.004 |
| 3 | −4.736 | −0.692 | −0.233 | 0.0 | −0.824 | −0.223 | −0.100 | −0.006 |
| 4 | −4.720 | −0.920 | −0.247 | 0.0 | −0.810 | −0.294 | −0.106 | −0.008 |
| 5 | −4.701 | −1.146 | −0.265 | 0.0 | −0.791 | −0.364 | −0.114 | −0.011 |
| 6 | −4.678 | −1.369 | −0.286 | 0.0 | −0.768 | −0.430 | −0.123 | −0.015 |
| 7 | −4.652 | −1.589 | −0.312 | 0.0 | −0.742 | −0.494 | −0.134 | −0.019 |
| 8 | −4.622 | −1.806 | −0.340 | 0.0 | −0.713 | −0.554 | −0.146 | −0.024 |
| 9 | −4.589 | −2.018 | −0.373 | 0.0 | −0.681 | −0.610 | −0.159 | −0.029 |
| 10 | −4.553 | −2.226 | −0.408 | 0.0 | −0.647 | −0.662 | −0.173 | −0.036 |
| 11 | −4.516 | −2.429 | −0.447 | 0.0 | −0.610 | −0.710 | −0.187 | −0.044 |
| 12 | −4.476 | −2.626 | −0.488 | 0.0 | −0.572 | −0.753 | −0.201 | −0.052 |
| 13 | −4.434 | −2.818 | −0.531 | 0.0 | −0.532 | −0.792 | −0.216 | −0.062 |
| 14 | −4.392 | −3.004 | −0.577 | 0.0 | −0.490 | −0.826 | −0.231 | −0.073 |
| 15 | −4.348 | −3.184 | −0.625 | 0.0 | −0.448 | −0.855 | −0.245 | −0.084 |
| 16 | −4.304 | −3.358 | −0.674 | 0.0 | −0.406 | −0.880 | −0.259 | −0.097 |
| 17 | −4.260 | −3.524 | −0.725 | 0.0 | −0.363 | −0.901 | −0.273 | −0.111 |
| 18 | −4.216 | −3.684 | −0.777 | 0.0 | −0.321 | −0.917 | −0.286 | −0.126 |
| 19 | −4.173 | −3.837 | −0.829 | 0.0 | −0.279 | −0.930 | −0.297 | −0.142 |
| 20 | −4.130 | −3.983 | −0.882 | 0.0 | −0.237 | −0.939 | −0.308 | −0.158 |

To obtain the information in single-particle channel, we adopt quantum cluster theories including cluster perturbation theory [77] and variational cluster approach [51], which have been widely applied to the studies of strongly correlated systems [52–60]. According to CPT and VCA, the interacting single-particle Green's function is expressed as

$$G^{-1}(\boldsymbol{k},\omega) = G_c^{-1}(\omega) - V(\boldsymbol{k}), \tag{B2}$$

where the infinite lattice is decomposed into identical clusters with finite size [see Fig. 2(a)], $G_c(\omega)$ is the exact single-particle Green's function of small clusters that can be obtain by ED, and $V(\boldsymbol{k})$ is the hoppings between clusters.

CPT is exempt from spontaneous symmetry breaking (SSB). To deal with the possible SSB, VCA can be employed by adding the corresponding test fields to the original system based on the variational principle. An SSB phase is obtained when the thermodynamic grand potential $\Omega$ is stationary at nonzero test fields. The corresponding order parameter $\bar{O} = \sum_{ab} O_{ab}\langle c_a^\dagger c_b\rangle$ can be computed by

$$\bar{O} = \frac{1}{N}\sum_{\boldsymbol{k}}\int_C \frac{d\omega}{2i\pi}\mathrm{Tr}[O(\boldsymbol{k})G(\boldsymbol{k},\omega)], \tag{B3}$$

where $C$ is the closed path that envelops the entire negative real axis, $O(\boldsymbol{k})$ is the Fourier transformation of $O_{ab}$, and $G(\boldsymbol{k},\omega)$ is obtained from Eq. (B2) at the stationary point. We calculate the 120° Néel order parameter via VCA with the orientations $\boldsymbol{e}_{1,2,3}$ of the 120° Néel moment shown in Fig. 2(a). The momentum distribution function $n(\boldsymbol{k})$ can be similarly obtained when $O_{ab}$ is taken as the identity matrix.

## APPENDIX C: FINITE-TEMPERATURE GREEN'S FUNCTION

It is hard to obtain the finite-temperature dynamical Green's function $A(\boldsymbol{k},\omega)_\beta$ on the 12-site cluster with ED due to the expensive costs. Here, we change the cluster solver from ED to TDVP to calculate the cluster Green's function $G_c(\omega)$ in Eq. (B2). TDVP [78,79] is a kind of state-of-the-art matrix product state (MPS) time-evolution method [80], which can evolve a state $|\psi\rangle$ by real-time (or imaginary-time) evolution slips $e^{-iHdt}$ (or $e^{-Hd\tau}$) to the state $e^{-iHt}|\psi\rangle$ (or $e^{-\beta H}|\psi\rangle$). Currently, mainstream open-source tensor libraries such as ITENSOR [81], MPSKIT [82], and TENPY [83] all support this technique.

At finite temperatures, dynamical single-particle correlation function can be written as

$$\langle c_i^\dagger(t)c_j\rangle_\beta = \frac{\mathrm{Tr}[e^{-\beta H}c_i^\dagger(t)c_j]}{Z} = \frac{\mathrm{Tr}(e^{-\beta H/2}e^{iHt}c_i^\dagger e^{-iHt}c_j e^{-\beta H/2})}{Z}. \tag{C1}$$

Here, $Z = \mathrm{Tr}(\rho)$, where $\rho = e^{-\beta H}$ is the density matrix operator. In the language of tensor, $\rho$ is a matrix product operator (MPO) in the given Hilbert space $\mathcal{H}$, which can be seen as a map in the Hilbert space $\rho: \mathcal{H}\to\mathcal{H}$ as shown in Fig. 12(a). We can convert the MPO $\rho = e^{-\beta H}$ to a super MPS $|\rho\rangle\rangle$, which is a vector belonging to the tensor product space $\mathcal{H}\otimes\mathcal{H}^*$. Hence, we can adopt the time-evolution MPS method like TDVP to obtain $|\rho\rangle\rangle$ by evolving the infinite-temperature density matrix $I\otimes I$, namely, an identity operator, with imaginary-time evolution $e^{-Hd\tau}\otimes I$ slice by slice to a given temperature [see Fig. 12(a)]. After obtaining $|\rho'\rangle\rangle$ ($\rho' = e^{\beta H/2}$), we can apply the annihilation operator $c_j\otimes I$ on $|\rho'\rangle\rangle$ and then use TDVP to evolve $c_j\otimes I|\rho'\rangle\rangle$ with real-time evolution $e^{-iHt}\otimes I$. When the right vector $(e^{-iHt}\otimes I)(c_j\otimes I)|\rho'\rangle\rangle$ and left vector $[(c_i\otimes I)(e^{-iHt}\otimes I)|\rho'\rangle\rangle]^\dagger$ are prepared, we can obtain the cluster Green's function $G_c(t)$ by

$$\langle c_i^\dagger(t)c_j\rangle_\beta = \langle\langle e^{-\beta H/2}e^{iHt}c_i^\dagger|e^{-iHt}c_j e^{-\beta H/2}\rangle\rangle/Z, \tag{C2}$$

which is illustrated in Fig. 12(b). Therefore, the thermodynamic average value in Eq. (C1) can be expressed by the dot product of the pure states in space $\mathcal{H}\otimes\mathcal{H}^*$ in Eq. (C2). After that, the frequency space cluster Green's function $G_c(\omega)$ can be obtained by the Fourier transform $G_c(\omega) = \int_0^{T_{\max}}\langle c_i^\dagger(t)c_j\rangle_\beta e^{i\omega t}dt$.

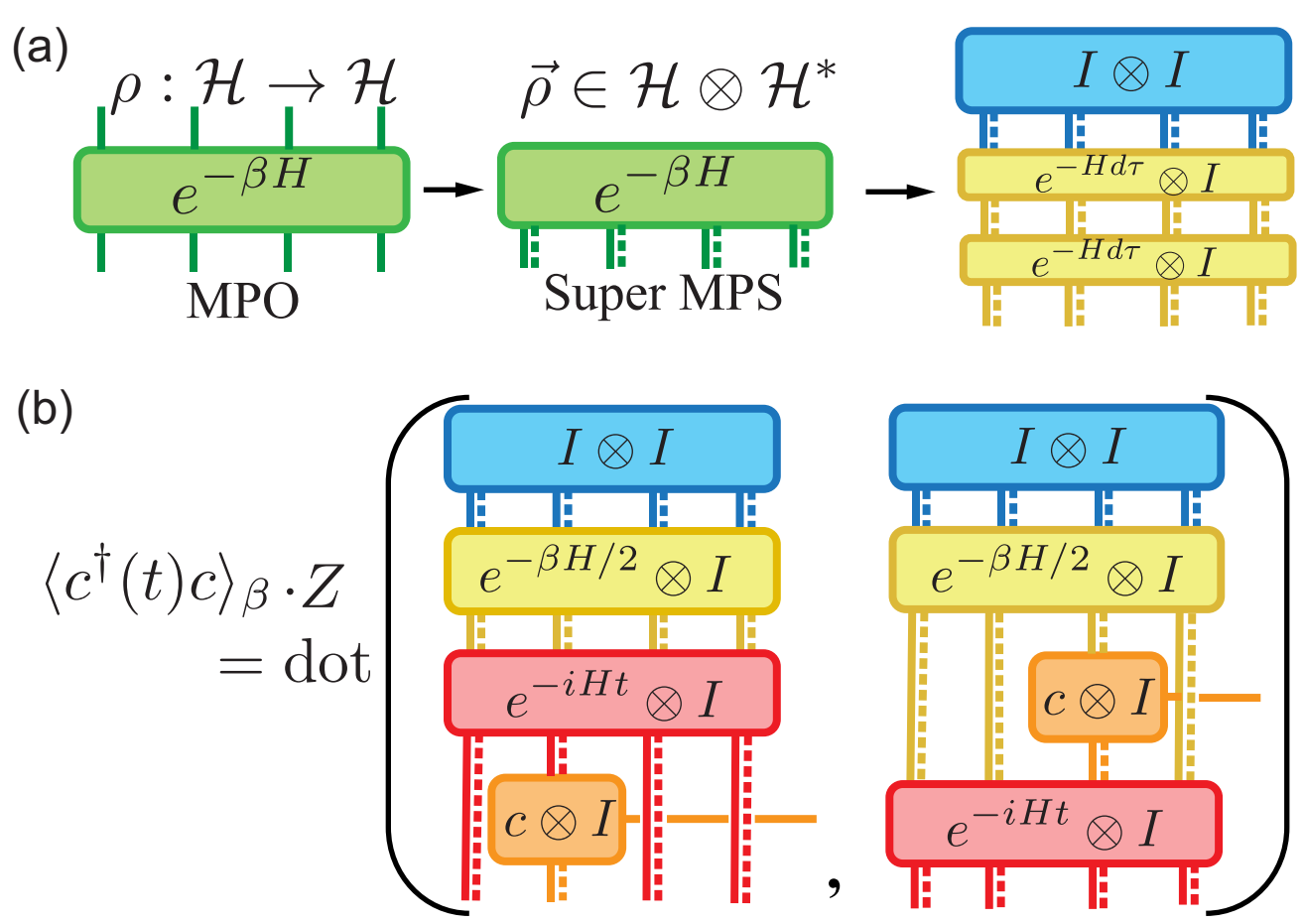

FIG. 12. Schematic illustration of the dynamical correlation functions at finite temperatures. (a) Relationship between MPO and super MPS, and the imaginary time evolution of super MPS from the infinite temperature. (b) The thermodynamic average value in Hilbert space $\mathcal{H}$ can be expressed as a dot product of pure states in a tensor product space $\mathcal{H}\otimes\mathcal{H}^*$.

Here, we adopt tensor libraries TENSORKIT and MPSKIT [82,84] to perform TDVP on the 12-site cluster. In practice, we take $\Delta\tau = 0.02$ and $\beta_{\max} = 0.68$ to evolve imaginary time. In order to obtain data with nearly equal temperature intervals, we pick the representative points [0.16, 0.18, 0.2, 0.22, 0.24, 0.28, 0.32, 0.4, 0.5, 0.68] of the imaginary-time series to perform real-time evolution. In real-time evolution, we set $\Delta t = 0.1$ and $T_{\max} = 50$, and put the max bond dimension up to $D = 512$ with $U(1)_{\mathrm{charge}} \times U(1)_{\mathrm{spin}}$ symmetry. The choices of $\Delta\tau$ and $\Delta t$ both meet the convergence conditions of TDVP in MPSKIT.JL. It is also noted that $2\pi/\Delta t$ can cover the relevant energy window. The selection of $D$ makes the truncation error of TDVP at the level of $10^{-2}$–$10^{-3}$ magnitude. The implementation details of the TDVP CPT method can be found in Refs. [76,85].

## APPENDIX D: CURRENT-CURRENT CORRELATION FUNCTION

The conductivity $\sigma_{\mathrm{dc}}$ also can be obtained by current-current correlation function, which is expressed as [86–88]

$$\sigma_{\mathrm{dc}} = \frac{\beta^2}{\pi} \Lambda_{xx}\left(\boldsymbol{k} = 0, \tau = \frac{\beta}{2}\right), \tag{D1}$$

where $\beta = 1/T, \Lambda_{xx} = \langle j_x(\tau) j_x(0) \rangle$ denotes the current-current correlation function, and $j_x$ is the current operator. We employ DQMC simulations for the moiré Hubbard model on an $8 \times 8$ triangular lattice with the software library [89] as a double check to determine the behaviors of resistivity.

Through systematic calculations of $\rho = 1/\sigma_{\mathrm{dc}}$ in the range of $10 < V^z < 50$ meV, we identify the transition from insulating to strange metallic and finally to Fermi liquid phases. Here, $U \simeq 17$ meV is selected to partially compensate finite-size effect induced energy gaps. Representative results at $V^z = 11$, 18, and 50 meV are marked by red, orange, and blue dots in Fig. 13, respectively. The corresponding color curves represent the fitting functions of $0.0026 e^{44.0332/T} + 1.93318$, $0.2620T + 0.0095$, and $0.0020T^2 + 0.1489$. In Fig. 13, the system starts with an insulating behavior characterized by $\rho \propto e^{1/T}$ (red curve) at low temperatures, transitions to a strange metal with $T$-linear resistivity $\rho \propto T$ (orange curve) extrapolating down to $T = 0$, and ultimately evolves into a Fermi liquid phase following $\rho \propto T^2$ (blue curve). Compared with the results in the main text, although the parameter regimes of the phase transitions are different, which might be due to the finite-size effect, the behavior of the resistivity evolution is qualitatively consistent.

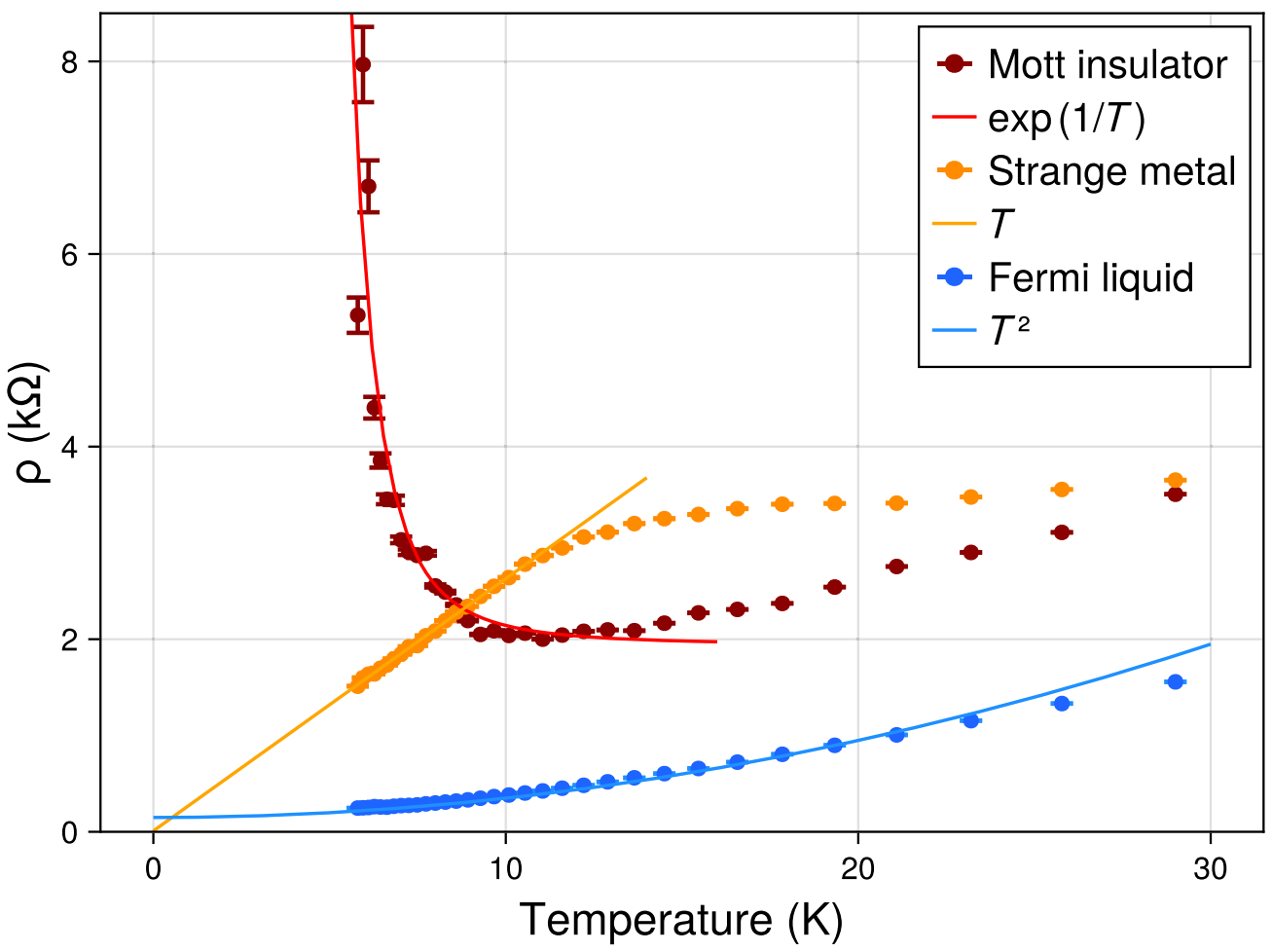

FIG. 13. DQMC results of resistivity versus temperature, revealing the insulator–strange metal–Fermi liquid transitions. The solid curves are the fitting functions. For each simulation, we run $10^6$ sweeps and measure $\Lambda_{xx}$ every 5 sweeps.

[1] J. M. Luttinger and J. C. Ward, *Ground-state energy of a many-fermion system. II*, Phys. Rev. **118**, 1417 (1960).
[2] J. M. Luttinger, *Fermi surface and some simple equilibrium properties of a system of interacting fermions*, Phys. Rev. **119**, 1153 (1960).
[3] M. Yamanaka, M. Oshikawa, and I. Affleck, *Nonperturbative approach to Luttinger's theorem in one dimension*, Phys. Rev. Lett. **79**, 1110 (1997).
[4] I. Dzyaloshinskii, *Some consequences of the Luttinger theorem: The Luttinger surfaces in non-Fermi liquids and Mott insulators*, Phys. Rev. B **68**, 085113 (2003).
[5] H. Ding, T. Yokoya, J. C. Campuzano, T. Takahashi, M. Randeria, M. R. Norman, T. Mochiku, K. Kadowaki, and J. Giapintzakis, *Spectroscopic evidence for a pseudogap in the normal state of underdoped high-$T_c$ superconductors*, Nature (London) **382**, 51 (1996).
[6] M. R. Norman, H. Ding, M. Randeria, J. C. Campuzano, T. Yokoya, T. Takeuchi, T. Takahashi, T. Mochiku, K. Kadowaki, P. Guptasarma, and D. G. Hinks, *Destruction of the Fermi surface in underdoped high-$T_c$ superconductors*, Nature (London) **392**, 157 (1998).
[7] A. Kaminski, S. Rosenkranz, H. M. Fretwell, J. C. Campuzano, Z. Li, H. Raffy, W. G. Cullen, H. You, C. G. Olson, C. M. Varma, and H. Höchst, *Spontaneous breaking of time-reversal symmetry in the pseudogap state of a high-$T_c$ superconductor*, Nature (London) **416**, 610 (2002).
[8] A. Damascelli, Z. Hussain, and Z.-X. Shen, *Angle-resolved photoemission studies of the cuprate superconductors*, Rev. Mod. Phys. **75**, 473 (2003).
[9] K. M. Shen, F. Ronning, D. H. Lu, F. Baumberger, N. J. C. Ingle, W. S. Lee, W. Meevasana, Y. Kohsaka, M. Azuma, M. Takano, H. Takagi, and Z.-X. Shen, *Nodal quasiparticles and antinodal charge ordering in* $Ca_{2-x}Na_xCuO_2Cl_2$, Science **307**, 901 (2005).
[10] A. Kanigel, M. R. Norman, M. Randeria, U. Chatterjee, S. Souma, A. Kaminski, H. M. Fretwell, S. Rosenkranz, M. Shi, T. Sato, T. Takahashi, Z. Z. Li, H. Raffy, K. Kadowaki, D. Hinks, L. Ozyuzer, and J. C. Campuzano, *Evolution of the pseudogap from Fermi arcs to the nodal liquid*, Nat. Phys. **2**, 447 (2006).
[11] R. Daou, N. Doiron-Leyraud, D. LeBoeuf, S. Y. Li, F. Laliberté, O. Cyr-Choinière, Y. J. Jo, L. Balicas, J. Q. Yan, J. S. Zhou, J. B. Goodenough, and L. Taillefer, *Linear temperature dependence of resistivity and change in the Fermi surface at the pseudogap critical point of a high-$T_c$ superconductor*, Nat. Phys. **5**, 31 (2009).
[12] A. V. Puchkov, P. Fournier, D. N. Basov, T. Timusk, A. Kapitulnik, and N. N. Kolesnikov, *Evolution of the*

*pseudogap state of high-$T_c$ superconductors with doping*, Phys. Rev. Lett. **77**, 3212 (1996).

[13] R. Daou, J. Chang, D. LeBoeuf, O. Cyr-Choinière, F. Laliberté, N. Doiron-Leyraud, B. J. Ramshaw, R. Liang, D. A. Bonn, W. N. Hardy, and L. Taillefer, *Broken rotational symmetry in the pseudogap phase of a high-$T_c$ superconductor*, Nature (London) **463**, 519 (2010).

[14] Y. Li, V. Balédent, G. Yu, N. Barišić, K. Hradil, R. A. Mole, Y. Sidis, P. Steffens, X. Zhao, P. Bourges, and M. Greven, *Hidden magnetic excitation in the pseudogap phase of a high-$T_c$ superconductor*, Nature (London) **468**, 283 (2010).

[15] F. Rullier-Albenque, H. Alloul, and G. Rikken, *High-field studies of superconducting fluctuations in high-$T_c$ cuprates: Evidence for a small gap distinct from the large pseudogap*, Phys. Rev. B **84**, 014522 (2011).

[16] S. I. Mirzaei, D. Stricker, J. N. Hancock, C. Berthod, A. Georges, E. van Heumen, M. K. Chan, X. Zhao, Y. Li, M. Greven, N. Barišić, and D. van der Marel, *Spectroscopic evidence for Fermi liquid-like energy and temperature dependence of the relaxation rate in the pseudogap phase of the cuprates*, Proc. Natl. Acad. Sci. U.S.A. **110**, 5774 (2013).

[17] K. B. Efetov, H. Meier, and C. Pépin, *Pseudogap state near a quantum critical point*, Nat. Phys. **9**, 442 (2013).

[18] S. Badoux, W. Tabis, F. Laliberté, G. Grissonnanche, B. Vignolle, D. Vignolles, J. Béard, D. A. Bonn, W. N. Hardy, R. Liang, N. Doiron-Leyraud, L. Taillefer, and C. Proust, *Change of carrier density at the pseudogap critical point of a cuprate superconductor*, Nature (London) **531**, 210 (2016).

[19] M. Mitrano, A. A. Husain, S. Vig, A. Kogar, M. S. Rak, S. I. Rubeck, J. Schmalian, B. Uchoa, J. Schneeloch, R. Zhong, G. D. Gu, and P. Abbamonte, *Anomalous density fluctuations in a strange metal*, Proc. Natl. Acad. Sci. U.S.A. **115**, 5392 (2018).

[20] R. L. Greene, P. R. Mandal, N. R. Poniatowski, and T. Sarkar, *The strange metal state of the electron-doped cuprates*, Annu. Rev. Condens. Matter Phys. **11**, 213 (2020).

[21] J. Ayres, M. Berben, M. Čulo, Y. T. Hsu, E. van Heumen, Y. Huang, J. Zaanen, T. Kondo, T. Takeuchi, J. R. Cooper, C. Putzke, S. Friedemann, A. Carrington, and N. E. Hussey, *Incoherent transport across the strange-metal regime of overdoped cuprates*, Nature (London) **595**, 661 (2021).

[22] S. Cai, J. Zhao, N. Ni, J. Guo, R. Yang, P. Wang, J. Han, S. Long, Y. Zhou, Q. Wu, X. Qiu, T. Xiang, R. J. Cava, and L. Sun, *The breakdown of both strange metal and superconducting states at a pressure-induced quantum critical point in iron-pnictide superconductors*, Nat. Commun. **14**, 3116 (2023).

[23] T. Kondo, R. Khasanov, T. Takeuchi, J. Schmalian, and A. Kaminski, *Competition between the pseudogap and superconductivity in the high-$T_c$ copper oxides*, Nature (London) **457**, 296 (2009).

[24] B. Keimer, S. A. Kivelson, M. R. Norman, S. Uchida, and J. Zaanen, *From quantum matter to high-temperature superconductivity in copper oxides*, Nature (London) **518**, 179 (2015).

[25] J. Chang, E. Blackburn, A. T. Holmes, N. B. Christensen, J. Larsen, J. Mesot, R. Liang, D. A. Bonn, W. N. Hardy, A. Watenphul, M. v. Zimmermann, E. M. Forgan, and S. M. Hayden, *Direct observation of competition between superconductivity and charge density wave order in* $YBa_2Cu_3O_{6.67}$, Nat. Phys. **8**, 871 (2012).

[26] P. W. Phillips, N. E. Hussey, and P. Abbamonte, *Stranger than metals*, Science **377**, eabh4273 (2022).

[27] J. M. Williams, A. J. Schultz, U. Geiser, K. D. Carlson, A. M. Kini, H. G. Wang, W.-K. Kwok, M.-H. Whangbo, and J. E. Schirber, *Organic superconductors—New benchmarks*, Science **252**, 1501 (1991).

[28] T. Katsufuji, Y. Taguchi, and Y. Tokura, *Transport and magnetic properties of a Mott-Hubbard system whose bandwidth and band filling are both controllable:* $R_{1-x}Ca_xTiO_{3+y/2}$, Phys. Rev. B **56**, 10145 (1997).

[29] K. Kanoda and R. Kato, *Mott physics in organic conductors with triangular lattices*, Annu. Rev. Condens. Matter Phys. **2**, 167 (2011).

[30] D. Faltermeier, J. Barz, M. Dumm, M. Dressel, N. Drichko, B. Petrov, V. Semkin, R. Vlasova, C. Meźière, and P. Batail, *Bandwidth-controlled Mott transition in* $\kappa - (BEDT-TTF)_2Cu[N(CN)_2]Br_xCl_{1-x}$*: Optical studies of localized charge excitations*, Phys. Rev. B **76**, 165113 (2007).

[31] M. Dumm, D. Faltermeier, N. Drichko, M. Dressel, C. Mézière, and P. Batail, *Bandwidth-controlled Mott transition in* $\kappa - (BEDT-TTF)_2Cu[N(CN)_2]Br_xCl_{1-x}$*: Optical studies of correlated carriers*, Phys. Rev. B **79**, 195106 (2009).

[32] R. Ang, Y. Miyata, E. Ieki, K. Nakayama, T. Sato, Y. Liu, W. J. Lu, Y. P. Sun, and T. Takahashi, *Superconductivity and bandwidth-controlled Mott metal-insulator transition in* $1T$-$TaS_{2-x}Se_x$, Phys. Rev. B **88**, 115145 (2013).

[33] H. C. Xu, Y. Zhang, M. Xu, R. Peng, X. P. Shen, V. N. Strocov, M. Shi, M. Kobayashi, T. Schmitt, B. P. Xie, and D. L. Feng, *Direct observation of the bandwidth control Mott transition in the* $NiS_{2-x}Se_x$ *multiband system*, Phys. Rev. Lett. **112**, 087603 (2014).

[34] L. Wang, E.-M. Shih, A. Ghiotto, L. Xian, D. A. Rhodes, C. Tan, M. Claassen, D. M. Kennes, Y. Bai, B. Kim, K. Watanabe, T. Taniguchi, X. Zhu, J. Hone, A. Rubio, A. N. Pasupathy, and C. R. Dean, *Correlated electronic phases in twisted bilayer transition metal dichalcogenides*, Nat. Mater. **19**, 861 (2020).

[35] Z. Zhang, Y. Wang, K. Watanabe, T. Taniguchi, K. Ueno, E. Tutuc, and B. J. LeRoy, *Flat bands in twisted bilayer transition metal dichalcogenides*, Nat. Phys. **16**, 1093 (2020).

[36] A. Ghiotto, E.-M. Shih, G. S. S. G. Pereira, D. A. Rhodes, B. Kim, J. Zang, A. J. Millis, K. Watanabe, T. Taniguchi, J. C. Hone, L. Wang, C. R. Dean, and A. N. Pasupathy, *Quantum criticality in twisted transition metal dichalcogenides*, Nature (London) **597**, 345 (2021).

[37] Y. Tang, L. Li, T. Li, Y. Xu, S. Liu, K. Barmak, K. Watanabe, T. Taniguchi, A. H. MacDonald, J. Shan, and K. F. Mak, *Simulation of Hubbard model physics in* $WSe_2/WS_2$ *moiré superlattices*, Nature (London) **579**, 353 (2020).

[38] T. Li, S. Jiang, L. Li, Y. Zhang, K. Kang, J. Zhu, K. Watanabe, T. Taniguchi, D. Chowdhury, L. Fu, J. Shan, and K. F. Mak, *Continuous Mott transition in semiconductor moiré superlattices*, Nature (London) **597**, 350 (2021).

[39] F. Wu, T. Lovorn, E. Tutuc, and A. H. MacDonald, *Hubbard model physics in transition metal dichalcogenide moiré bands*, Phys. Rev. Lett. **121**, 026402 (2018).

[40] T. Devakul, V. Crépel, Y. Zhang, and L. Fu, *Magic in twisted transition metal dichalcogenide bilayers*, Nat. Commun. **12**, 6730 (2021).

[41] J. Zang, J. Wang, J. Cano, A. Georges, and A. J. Millis, *Dynamical mean-field theory of moiré bilayer transition metal dichalcogenides: Phase diagram, resistivity, and quantum criticality*, Phys. Rev. X **12**, 021064 (2022).

[42] K.-Y. Yang, T. M. Rice, and F.-C. Zhang, *Phenomenological theory of the pseudogap state*, Phys. Rev. B **73**, 174501 (2006).

[43] T. D. Stanescu and G. Kotliar, *Fermi arcs and hidden zeros of the Green function in the pseudogap state*, Phys. Rev. B **74**, 125110 (2006).

[44] S. Sakai, Y. Motome, and M. Imada, *Evolution of electronic structure of doped Mott insulators: Reconstruction of poles and zeros of Green's function*, Phys. Rev. Lett. **102**, 056404 (2009).

[45] S. Sakai, Y. Motome, and M. Imada, *Doped high-$T_c$ cuprate superconductors elucidated in the light of zeros and poles of the electronic Green's function*, Phys. Rev. B **82**, 134505 (2010).

[46] N. Lin, E. Gull, and A. J. Millis, *Physics of the pseudogap in eight-site cluster dynamical mean-field theory: Photoemission, Raman scattering, and in-plane and c-axis conductivity*, Phys. Rev. B **82**, 045104 (2010).

[47] M. Bélanger, J. Fournier, and D. Sénéchal, *Superconductivity in the twisted bilayer transition metal dichalcogenide* $WSe_2$*: A quantum cluster study*, Phys. Rev. B **106**, 235135 (2022).

[48] H. Pan, F. Wu, and S. Das Sarma, *Band topology, Hubbard model, Heisenberg model, and Dzyaloshinskii-Moriya interaction in twisted bilayer* $WSe_2$, Phys. Rev. Res. **2**, 033087 (2020).

[49] F. Wu, T. Lovorn, E. Tutuc, I. Martin, and A. H. MacDonald, *Topological insulators in twisted transition metal dichalcogenide homobilayers*, Phys. Rev. Lett. **122**, 086402 (2019).

[50] D. Sénéchal, D. Perez, and D. Plouffe, *Cluster perturbation theory for Hubbard models*, Phys. Rev. B **66**, 075129 (2002).

[51] M. Potthoff, M. Aichhorn, and C. Dahnken, *Variational cluster approach to correlated electron systems in low dimensions*, Phys. Rev. Lett. **91**, 206402 (2003).

[52] D. Sénéchal, P.-L. Lavertu, M.-A. Marois, and A.-M. S. Tremblay, *Competition between antiferromagnetism and superconductivity in high-$T_c$ cuprates*, Phys. Rev. Lett. **94**, 156404 (2005).

[53] P. Sahebsara and D. Sénéchal, *Hubbard model on the triangular lattice: Spiral order and spin liquid*, Phys. Rev. Lett. **100**, 136402 (2008).

[54] S.-L. Yu, X. C. Xie, and J.-X. Li, *Mott physics and topological phase transition in correlated Dirac fermions*, Phys. Rev. Lett. **107**, 010401 (2011).

[55] S.-L. Yu and J.-X. Li, *Chiral superconducting phase and chiral spin-density-wave phase in a Hubbard model on the kagome lattice*, Phys. Rev. B **85**, 144402 (2012).

[56] M. Kohno, *Mott transition in the two-dimensional Hubbard model*, Phys. Rev. Lett. **108**, 076401 (2012).

[57] S. Rachel, M. Laubach, J. Reuther, and R. Thomale, *Quantum paramagnet in a $\pi$ flux triangular lattice Hubbard model*, Phys. Rev. Lett. **114**, 167201 (2015).

[58] Z.-L. Gu, K. Li, and J.-X. Li, *Quantum cluster approach to the topological invariants in correlated Chern insulators*, New J. Phys. **21**, 073016 (2019).

[59] Z. Chen, Y. Wang, S. N. Rebec, T. Jia, M. Hashimoto, D. Lu, B. Moritz, R. G. Moore, T. P. Devereaux, and Z.-X. Shen, *Anomalously strong near-neighbor attraction in doped 1D cuprate chains*, Science **373**, 1235 (2021).

[60] C. Gu, Z.-L. Gu, S.-L. Yu, and J.-X. Li, *Spectral evolution of the $s = \frac{1}{2}$ antiferromagnetic Heisenberg model: From one to two dimensions*, Phys. Rev. B **108**, 224418 (2023).

[61] T. Timusk and B. Statt, *The pseudogap in high-temperature superconductors: An experimental survey*, Rep. Prog. Phys. **62**, 61 (1999).

[62] V. M. Loktev, R. M. Quick, and S. G. Sharapov, *Phase fluctuations and pseudogap phenomena*, Phys. Rep. **349**, 1 (2001).

[63] D. Sénéchal and A.-M. S. Tremblay, *Hot spots and pseudogaps for hole- and electron-doped high-temperature superconductors*, Phys. Rev. Lett. **92**, 126401 (2004).

[64] E. Gull, O. Parcollet, and A. J. Millis, *Superconductivity and the pseudogap in the two-dimensional Hubbard model*, Phys. Rev. Lett. **110**, 216405 (2013).

[65] T. Valla, A. V. Fedorov, J. Lee, J. C. Davis, and G. D. Gu, *The ground state of the pseudogap in cuprate superconductors*, Science **314**, 1914 (2006).

[66] L. Landau and E. Lifshitz, *Statistical Physics: Theory of the Condensed State*, Course of Theoretical Physics (Elsevier Science, New York, 1980).

[67] P. Coleman, *Introduction to Many-Body Physics* (Cambridge University Press, Cambridge, England, 2015).

[68] T. D. Stanescu, P. Phillips, and T.-P. Choy, *Theory of the Luttinger surface in doped Mott insulators*, Phys. Rev. B **75**, 104503 (2007).

[69] P. Phillips, T.-P. Choy, and R. G. Leigh, *Mottness in high-temperature copper-oxide superconductors*, Rep. Prog. Phys. **72**, 036501 (2009).

[70] J. C. Slater, *Magnetic effects and the Hartree-Fock equation*, Phys. Rev. **82**, 538 (1951).

[71] Q.-H. Wang and D.-H. Lee, *Quasiparticle scattering interference in high-temperature superconductors*, Phys. Rev. B **67**, 020511(R) (2003).

[72] H.-Y. Zhang and J.-X. Li, *Quasiparticle scattering interference in iron pnictides: A probe of the origin of nematicity*, Phys. Rev. B **94**, 075153 (2016).

[73] D.-B. Zhang, Q. Han, and Z. D. Wang, *Local and global patterns in quasiparticle interference: A reduced response function approach*, Phys. Rev. B **100**, 205112 (2019).

[74] A. Georges, G. Kotliar, W. Krauth, and M. J. Rozenberg, *Dynamical mean-field theory of strongly correlated fermion systems and the limit of infinite dimensions*, Rev. Mod. Phys. **68**, 13 (1996).

[75] S. Kunisada, S. Isono, Y. Kohama, S. Sakai, C. Bareille, S. Sakuragi, R. Noguchi, K. Kurokawa, K. Kuroda, Y. Ishida, S. Adachi, R. Sekine, T. K. Kim, C. Cacho, S. Shin, T. Tohyama, K. Tokiwa, and T. Kondo, *Observation of small Fermi pockets protected by clean* $CuO_2$ *sheets of a high-$T_c$ superconductor*, Science **369**, 833 (2020).

[76] https://github.com/ZongYongyue/tWSe2-pseudogap.

[77] D. Sénéchal, D. Perez, and M. Pioro-Ladrière, *Spectral weight of the Hubbard model through cluster perturbation theory*, Phys. Rev. Lett. **84**, 522 (2000).

[78] J. Haegeman, J. I. Cirac, T. J. Osborne, I. Pižorn, H. Verschelde, and F. Verstraete, *Time-dependent variational principle for quantum lattices*, Phys. Rev. Lett. **107**, 070601 (2011).

[79] J. Haegeman, C. Lubich, I. Oseledets, B. Vandereycken, and F. Verstraete, *Unifying time evolution and optimization with matrix product states*, Phys. Rev. B **94**, 165116 (2016).

[80] S. Paeckel, T. Köhler, A. Swoboda, S. R. Manmana, U. Schollwöck, and C. Hubig, *Time-evolution methods for matrix-product states*, Ann. Phys. (Amsterdam) **411**, 167998 (2019).

[81] M. Fishman, S. R. White, and E. M. Stoudenmire, *The ITENSOR software library for tensor network calculations*, SciPost Phys. Codebases 4 (2022), 10.21468/SciPostPhysCodeb.4.

[82] M. Van Damme, L. Devos, and J. Haegeman, MPSKIT, 10.5281/zenodo.10654900 (2025).

[83] J. Hauschild *et al.*, *Tensor network PYTHON (TENPY) version 1*, SciPost Phys. Codebases **41**, 41 (2024), 10.21468/SciPostPhysCodeb.41.

[84] L. Devos and J. Haegeman, *TENSORKIT.JL: A Julia package for large-scale tensor computations, with a hint of category theory*, arXiv:2508.10076.

[85] https://github.com/ZongYongyue/DynamicalCorrelators.jl.

[86] D. J. Scalapino, S. R. White, and S. C. Zhang, *Superfluid density and the Drude weight of the Hubbard model*, Phys. Rev. Lett. **68**, 2830 (1992).

[87] D. J. Scalapino, S. R. White, and S. Zhang, *Insulator, metal, or superconductor: The criteria*, Phys. Rev. B **47**, 7995 (1993).

[88] N. Paris, K. Bouadim, F. Hebert, G. G. Batrouni, and R. T. Scalettar, *Quantum Monte Carlo study of an interaction-driven band-insulator–to–metal transition*, Phys. Rev. Lett. **98**, 046403 (2007).

[89] C. Bauer, *MONTECARLO.JL: Classical and quantum Monte Carlo simulations in Julia*, 10.5281/zenodo.3819449 (2020).